# Low-Earth Orbit Determination from Gravity Gradient Measurements

Xiucong Sun [a,b], Pei Chen [a], Christophe Macabiau [b], Chao Han [a,*]

[a] *School of Astronautics, Beihang University, Beijing 100191, China*

[b] *TELECOM Lab, Ecole Nationale de l'Aviation Civile (ENAC), Toulouse 31055, France*

**Abstract** An innovative orbit determination method which makes use of gravity gradients for Low-Earth-Orbiting satellites is proposed. The measurement principle of gravity gradiometry is briefly reviewed and the sources of measurement error are analyzed. An adaptive hybrid least squares batch filter based on linearization of the orbital equation and unscented transformation of the measurement equation is developed to estimate the orbital states and the measurement biases. The algorithm is tested with the actual flight data from the European Space Agency's Gravity field and steady-state Ocean Circulation Explorer (GOCE). The orbit determination results are compared with the GPS-derived orbits. The radial and cross-track position errors are on the order of tens of meters, whereas the along-track position error is over one order of magnitude larger. The gravity gradient based orbit determination method is promising for potential use in GPS-denied spacecraft navigation.

*Keywords*: Gravity gradient; Gravity gradiometry; LEO orbit determination; Adaptive hybrid least squares; GOCE

## 1. Introduction

Satellite orbit determination usually relies on geometric measurements. Two typical examples are the ground-based radar tracking and the Global Positioning System (GPS), both utilizing electromagnetic wave propagation to measure relative distance and direction. Current GPS technology achieves centimeter-level accuracy for Low Earth Orbiting (LEO) satellites with dual-frequency carrier phases [1,2]. An alternative method to the geometry-based orbit determination is the geophysical navigation, which derives position from local geophysical data. One representative is the magnetometer-based autonomous navigation. Orbital position errors ranging from a few to a hundred kilometers have been achieved with real flight data from several LEO satellites [3-6]. Despite the low accuracy, the geophysical navigation does not need any support from ground stations or any other satellites and is thus suitable for autonomous spacecraft operation in GPS-denied environments.

[*] Corresponding author. Tel.: +86 10 82339583.
E-mail address: hanchao@buaa.edu.cn (C. Han).

Besides the magnetic field, the gravity is another kind of geophysical information that can be exploited for orbit determination. The gravity field of the Earth is more stable than the magnetic field. The effects of the gravity field on satellites include two aspects. Firstly, the gravitational attraction is the main driving force of the orbital motion. The modeling of the gravity field is of crucial importance to precise orbit prediction. Secondly, the gravity gradients vary as functions of position and orientation and can be measured by a spaceborne gradiometer [7,8]. With a known gravity model as well as the satellite orientation information, the orbit could be estimated from the gravity gradient observations.

According to the gravitational potential theory, the gravity filed is usually described in terms of multipoles, which provide integral characteristics of the matter distribution inside an astronomical body. The multipolar expansion of gravity has been proven to be useful in celestial mechanics. For example, the long-term effects on satellite orbital motion due to lower-order zonal harmonics are well investigated for an arbitrary orientation of the rotation axis of the body [9-11]. The multipoles are crucial also in several satellite tests of fundamental physics such as the LARES/LAGEOS frame-dragging experiment [12-14].

Nowadays, the accuracy of the Earth's gravity model has been improved dramatically since the development of modern space-geodetic techniques such as GPS, VLBI (Very Long Baseline Interferometry), and SLR (Satellite Laser Ranging). The new developed Earth Gravitational Model 2008 (EGM2008) is complete to degree 2190 and order 2159 by combination of satellite geodetic data and high-resolution surface gravimetry [15]. More recently, general relativity has entered the field of geodesy towards better interpretation of high-precision geodetic measurements with a post-Newtonian formalism [16-18]. The endeavor has led to the adoption of a series of resolutions on relativistic reference systems and time scales by the International Astronomical Union (IAU). Vice versa, the space-geodetic measurements can also be used to explore the relativistic effects. Orbiting superconducting gravity gradiometers have been recently proposed to detect the gravitomagnetic field, which causes the "frame-dragging effect" or "Lense-Thirring effect" [19,20].

The application of gravity gradiometry for navigation has been studied since the 1960s. One of the main research interests is to incorporate a gradiometer into an airborne or shipborne inertial navigation system (INS) for real-time compensation of gravity model uncertainties [21-24]. Metzger and Jircitano [25] presented an early form of map-matching technique by cross-correlating the sensed gravity gradients with previously mapped values. The premise was to let a vehicle travel a course twice and to compute the state lag from gravity gradient measurements on both passes. Affleck and Jircitano [26] later developed a passive gravity gradiometer navigation system in which the gravity gradient map was not provided by a first flight but generated from the terrain elevation data base. An optimal filter was designed to update positions and to correct instrument errors. During the following twenty years, further contributions were made on this topic, including the Fast Fourier Transformation based rapid map generation [27], extended application to a hypersonic cruise [28], and feasibility investigation using a modern gradiometer, which is defined as a gravity gradiometer projected to be available within the next 10 years [29]. By contrast, little research has been conducted on the applications of gradiometry for spacecraft navigation. The major difference between the inertial navigation aiding and the application for spacecraft navigation lies in the fact that high-precision attitude information could be easily obtained onboard a satellite by star trackers. This decouples position estimation from attitude estimation. In addition, the higher frequency terrain contributions are dramatically attenuated at the height of a spacecraft. A truncated spherical harmonic gravity model will be accurate enough for space users.

In 2009, ESA's Gravity field and steady-state Ocean Circulation Explorer (GOCE) satellite was launched into a sun-synchronous LEO orbit to determine the Earth's gravity field [30]. The satellite carried an Electrostatic Gravity Gradiometer (EGG) and measured gravity gradients from an unprecedented low altitude of about 260 km in space. Post flight analysis showed that a noise density level of 0.01 $E/\sqrt{Hz}$ was achieved within the measurement bandwidth (MBW) from $5\times10^{-3}$ to 0.1 Hz [8]. The satellite was also equipped with three advanced star trackers and two dual-frequency GPS receivers. These conditions make GOCE an ideal testbed for the research of gravity gradient based space navigation. Sun et al. [31] introduced an idea of using full-tensor gravity gradients combined with high-precision attitudes to determine a spacecraft's position. A least squares searching algorithm was developed and a mean positioning error of 620 m was achieved with real GOCE data. An eigendecomposition method using the $J_2$ gravity model was presented in Chen et al. [32] and position errors ranging from 421 to 2690 m were achieved.

In the previous studies mentioned above, the gravity gradient observation errors are modeled as low-level white noise only. In fact, the gradiometer measurements contain significant biases and low-frequency noises. The present work deals with the biases and the drifts in the actual measurements. The noise characteristics are investigated and a simplified observation error model is formulated. An adaptive hybrid least squares batch filter is developed to estimate the orbital states and the biases. The filter combines the advantages of the linear approximation of the orbital equation and the unscented transformation of gravity gradient observations to achieve fast and accurate orbit determination. The measurement time span at each iteration step is adaptively adjusted to restrict the linearization errors and thus to guarantee convergence. An augmented state iterated least squares filter is implemented thereafter to further estimate the drifts. The algorithms are tested with real GOCE data and the orbit determination results are compared with the Precise Science Orbit (PSO) solutions derived from the GPS system.

The remainder of this paper is organized as follows. Section 2 briefly reviews the measurement principle of GOCE gravity gradiometry and investigates the sources of measurement error in gravity gradient retrieval. Section 3 presents the orbital dynamic model, the gravity gradient observation model, and the measurement error model for orbit determination. Section 4 summarizes the iterated least squares filter and the unscented least squares filter for nonlinear estimation and presents the algorithm of the adaptive hybrid least squares filter. Section 5 presents the orbit determination results obtained with real GOCE data. Conclusions are drawn in Section 6.

## 2. GOCE gravity gradiometry

A differential accelerometry technique was employed by GOCE to measure gravity gradients. The gradiometer was placed close to the spacecraft's center of mass (CoM) and consisted of three orthogonal pairs of capacitive accelerometers. Each accelerometer had two ultra-sensitive axes and one less-sensitive axis. The three pairs of accelerometers were mounted at the ends of three baselines having an approximate length of 0.5 m. The gradiometer reference frame (GRF) is materialized by the three orthogonal baselines with the $X$ axis in the flight direction, the $Y$ axis normal to the orbit plane, and the $Z$ axis radially downwards, as depicted in Fig. 1. Inside each accelerometer, a platinum-rhodium proof mass was electrostatically levitated at the center of a cage, leading to control voltages that were representative of the sum of the non-gravitational accelerations at the location of the proof mass [33]. The gravity gradients were contained in the accelerometer differences.

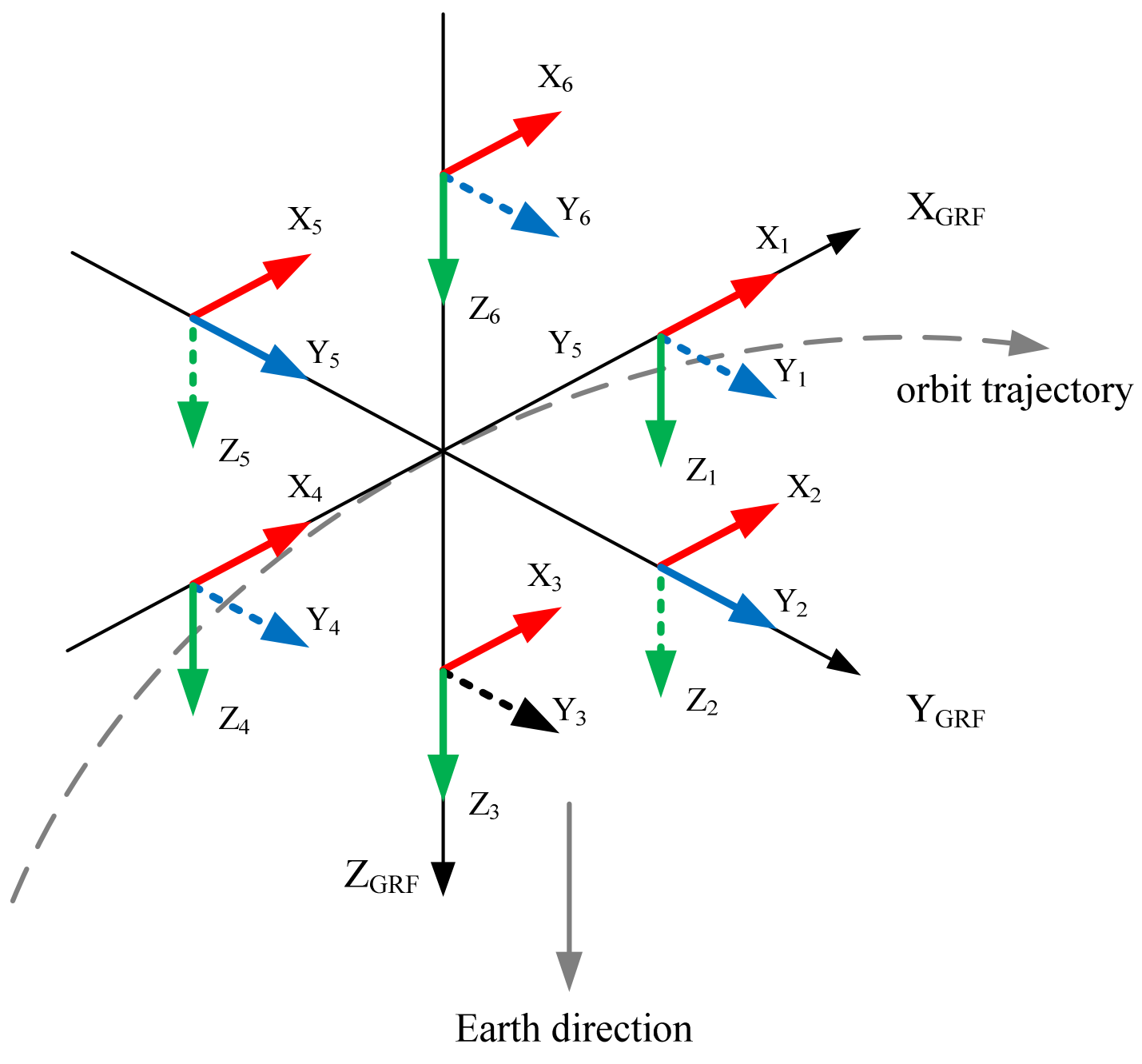

**Fig. 1.** The arrangement of the six accelerometers inside GOCE gradiometer and the orientation of the GRF reference frame. The solid arrows at each accelerometer show the ultra-sensitive axes and the dashed arrows show the less sensitive axes.

### *2.1. Measurement principle*

To describe the measurement process, an ideal gradiometer is considered by assuming that:

(i). The centers of the three baselines are coincident;

(ii). The baselines are mutually perpendicular and perfectly aligned with the three axes of GRF;

(iii). The accelerometers occupy their nominal positions and their axes are also aligned with GRF;

(iv). The accelerometers can sense the local non-gravitational accelerations exactly, which means that there are no errors in the conversion of control voltages.

Under these ideal conditions, the output of each accelerometer has the following expression

$$\boldsymbol{a}_i = \left(\boldsymbol{\Omega}^2 + \dot{\boldsymbol{\Omega}} - \boldsymbol{V}\right)\boldsymbol{r}_i + \boldsymbol{d}_a \tag{1}$$

where $i$ is the identifier of the accelerometer, $\boldsymbol{r}_i$ is the vector from CoM to the position of the $i$th accelerometer, $\boldsymbol{\Omega}^2\boldsymbol{r}_i$ is the centrifugal acceleration induced by the spacecraft's angular rotation, $\dot{\boldsymbol{\Omega}}\boldsymbol{r}_i$ is the Euler acceleration, $\boldsymbol{V}$ is the gravity gradient tensor (GGT), and $\boldsymbol{d}_a$ is the non-gravitational acceleration at CoM. The accelerations due to the higher order ( ≥ 3rd) derivatives of the gravitational potential, relative motion of the cage (e.g., Coriolis effects), self-gravity, and the coupling with the external magnetic field are not considered. The matrices $\boldsymbol{\Omega}$, $\dot{\boldsymbol{\Omega}}$, and $\boldsymbol{\Omega}^2$ are defined as

$$\boldsymbol{\Omega} = \begin{bmatrix} 0 & -\omega_z & \omega_y \\ \omega_z & 0 & -\omega_x \\ -\omega_y & \omega_x & 0 \end{bmatrix} \tag{2}$$

$$\dot{\boldsymbol{\Omega}} = \begin{bmatrix} 0 & -\dot{\omega}_z & \dot{\omega}_y \\ \dot{\omega}_z & 0 & -\dot{\omega}_x \\ -\dot{\omega}_y & \dot{\omega}_x & 0 \end{bmatrix} \tag{3}$$

$$\boldsymbol{\Omega}^2 = \begin{bmatrix} -\left(\omega_y^2 + \omega_z^2\right) & \omega_x\omega_y & \omega_x\omega_z \\ \omega_x\omega_y & -\left(\omega_x^2 + \omega_z^2\right) & \omega_y\omega_z \\ \omega_x\omega_z & \omega_y\omega_z & -\left(\omega_x^2 + \omega_y^2\right) \end{bmatrix} \tag{4}$$

where $\omega_x$, $\omega_y$, and $\omega_z$ are the angular velocities, and $\dot{\omega}_x$, $\dot{\omega}_y$, and $\dot{\omega}_z$ are the angular accelerations. All the vectors and matrices are expressed in the GRF frame.

The non-gravitational acceleration $\boldsymbol{d_a}$ is first isolated by forming the common-mode (CM) and the differential-mode (DM) accelerations

$$\boldsymbol{a}_{c,ij} = \frac{1}{2}\left(\boldsymbol{a}_i + \boldsymbol{a}_j\right) = \left(\boldsymbol{\Omega}^2 + \dot{\boldsymbol{\Omega}} - \boldsymbol{V}\right)\boldsymbol{c} + \boldsymbol{d_a} \tag{5}$$

$$\boldsymbol{a}_{d,ij} = \frac{1}{2}\left(\boldsymbol{a}_i - \boldsymbol{a}_j\right) = \frac{1}{2}\left(\boldsymbol{\Omega}^2 + \dot{\boldsymbol{\Omega}} - \boldsymbol{V}\right)\boldsymbol{L}_{ij} \tag{6}$$

where $ij \in \{14, 25, 36\}$ represents the index of the accelerometer pairs, $\boldsymbol{c}$ is the vector from CoM to the center of the baselines, and $\boldsymbol{L}_{ij}$ is the vector from the *j*th to the *i*th accelerometer.

Then combine the three DM accelerations to form a matrix equation

$$\boldsymbol{A}_d = \frac{1}{2}\left(\boldsymbol{\Omega}^2 + \dot{\boldsymbol{\Omega}} - \boldsymbol{V}\right)\boldsymbol{L} \tag{7}$$

with

$$\boldsymbol{A}_d = \begin{bmatrix} \boldsymbol{a}_{d,14} & \boldsymbol{a}_{d,25} & \boldsymbol{a}_{d,36} \end{bmatrix} \tag{8}$$

$$\boldsymbol{L} = \begin{bmatrix} \boldsymbol{L}_{14} & \boldsymbol{L}_{25} & \boldsymbol{L}_{36} \end{bmatrix} = \begin{bmatrix} L_x & 0 & 0 \\ 0 & L_y & 0 \\ 0 & 0 & L_z \end{bmatrix} \tag{9}$$

where $L_x$, $L_y$, and $L_z$ are the lengths of the three baselines. The right-hand side of Eq. (7) contains the GGT, the centrifugal accelerations, and the Euler accelerations. Based on the symmetry of $\boldsymbol{\Omega}^2$ and $\boldsymbol{V}$ and the skew-symmetry of $\dot{\boldsymbol{\Omega}}$, the Euler accelerations can be isolated

$$\boldsymbol{A}_d - \boldsymbol{A}_d^T = \dot{\boldsymbol{\Omega}}\boldsymbol{L} \tag{10}$$

$$\boldsymbol{A}_d + \boldsymbol{A}_d^T = \left(\boldsymbol{\Omega}^2 - \boldsymbol{V}\right)\boldsymbol{L} \tag{11}$$

The gravity gradient tensor is finally retrieved as follows

$$\boldsymbol{V} = -\left[\left(\boldsymbol{A}_d + \boldsymbol{A}_d^T\right)\boldsymbol{L}^{-1} - \boldsymbol{\Omega}^2\right] \tag{12}$$

*2.2. Error sources*

The ideal GGT retrieval procedure is subject to several sources of measurement error. For the most part, the errors are attributed to the accelerometers. First, accelerations are measured by voltage signals. The transformation from voltage to acceleration requires the accurate knowledge of the electrostatic gains and the read-out gain. Uncertainties of these gains result in scale factors in the outputs [33]. Second, non-linearity exists in the electronic components and the transfer functions of the control loop, leading to an additional quadratic term, which is proportional to the square of the input acceleration. Last but not least, the accelerometer outputs contain unknown biases and noises due to the intrinsic imperfection. In addition to the accelerometer-related errors, the gravity gradiometer also has geometric imperfections. For example, the misalignments of the six accelerometers induce small rotation angles with respect to the nominal GRF frame and non-perfect orthogonality between the accelerometer axes causes cross coupling errors.

Taking all of the factors above into consideration, the actual output of each accelerometer should be written as

$$\tilde{\boldsymbol{a}}_i = \left(\boldsymbol{K}_i + d\boldsymbol{R}_i + d\boldsymbol{S}_i\right)\boldsymbol{a}_i + \boldsymbol{K2}_i\boldsymbol{a}_i^2 + \boldsymbol{b}_i + \boldsymbol{n}_i \tag{13}$$

where $\boldsymbol{K}_i$ and $\boldsymbol{K2}_i$ are the scale factor and the quadratic factor, $d\boldsymbol{R}_i$ is the rotation matrix due to misalignments, $d\boldsymbol{S}_i$ is the accelerometer inter-axis coupling matrix, and $\boldsymbol{b}_i$ and $\boldsymbol{n}_i$ are the accelerometer bias and noise.

The quadratic factor of each accelerometer was identified and calibrated in flight and the calibration error is negligible [34]. Let $\boldsymbol{M}_i$ denote the sum of $\boldsymbol{K}_i$, $d\boldsymbol{R}_i$, and $d\boldsymbol{S}_i$. According to Eq. (6), the actual obtained CM and DM accelerations are

$$\begin{bmatrix} \tilde{\boldsymbol{a}}_{c,ij} \\ \tilde{\boldsymbol{a}}_{d,ij} \end{bmatrix} = \boldsymbol{M}_{ij} \begin{bmatrix} \boldsymbol{a}_{c,ij} \\ \boldsymbol{a}_{d,ij} \end{bmatrix} + \begin{bmatrix} \boldsymbol{b}_{c,ij} \\ \boldsymbol{b}_{d,ij} \end{bmatrix} + \begin{bmatrix} \boldsymbol{n}_{c,ij} \\ \boldsymbol{n}_{d,ij} \end{bmatrix} \tag{14}$$

with

$$\boldsymbol{M}_{ij} = \frac{1}{2} \begin{bmatrix} \boldsymbol{M}_i + \boldsymbol{M}_j & \boldsymbol{M}_i - \boldsymbol{M}_j \\ \boldsymbol{M}_i - \boldsymbol{M}_j & \boldsymbol{M}_i + \boldsymbol{M}_j \end{bmatrix} \tag{15}$$

$$\begin{bmatrix} \boldsymbol{b}_{c,ij} \\ \boldsymbol{b}_{d,ij} \end{bmatrix} = \frac{1}{2} \begin{bmatrix} \boldsymbol{b}_i + \boldsymbol{b}_j \\ \boldsymbol{b}_i - \boldsymbol{b}_j \end{bmatrix}, \quad \begin{bmatrix} \boldsymbol{n}_{c,ij} \\ \boldsymbol{n}_{d,ij} \end{bmatrix} = \frac{1}{2} \begin{bmatrix} \boldsymbol{n}_i + \boldsymbol{n}_j \\ \boldsymbol{n}_i - \boldsymbol{n}_j \end{bmatrix} \tag{16}$$

where $\boldsymbol{a}_{c,ij}$ and $\boldsymbol{a}_{d,ij}$ are true CM and DM accelerations, $\boldsymbol{M}_{ij}$ is a scaling calibration matrix related to $\boldsymbol{M}_i$ and $\boldsymbol{M}_j$, and $\boldsymbol{b}_{c,ij}$, $\boldsymbol{b}_{d,ij}$, $\boldsymbol{n}_{c,ij}$, and $\boldsymbol{n}_{d,ij}$ are CM and DM biases and noises, respectively. To obtain $\boldsymbol{a}_{c,ij}$ and $\boldsymbol{a}_{d,ij}$ from $\tilde{\boldsymbol{a}}_{c,ij}$ and $\tilde{\boldsymbol{a}}_{d,ij}$, the inverse of the calibration matrix (ICM) is needed. The ICMs of the three baselines were directly determined from an in-flight satellite shaking procedure, which was achieved by the ion thruster and the gradiometer calibration device. Details of the calibration are found in [34]. Let $\hat{\boldsymbol{M}}_{ij}^{-1}$ and $\Delta\boldsymbol{M}_{ij}^{-1}$ denote the calibration value and the calibration error of $\boldsymbol{M}_{ij}^{-1}$, respectively. The CM and DM accelerations after ICM calibration will be

$$\begin{bmatrix} \hat{\boldsymbol{a}}_{c,ij} \\ \hat{\boldsymbol{a}}_{d,ij} \end{bmatrix} = \hat{\boldsymbol{M}}_{ij}^{-1} \begin{bmatrix} \tilde{\boldsymbol{a}}_{c,ij} \\ \tilde{\boldsymbol{a}}_{d,ij} \end{bmatrix} \cong \begin{bmatrix} \boldsymbol{a}_{c,ij} \\ \boldsymbol{a}_{d,ij} \end{bmatrix} + \Delta\hat{\boldsymbol{M}}_{ij}^{-1} \begin{bmatrix} \boldsymbol{a}_{c,ij} \\ \boldsymbol{a}_{d,ij} \end{bmatrix} + \hat{\boldsymbol{M}}_{ij}^{-1} \begin{bmatrix} \boldsymbol{b}_{c,ij} + \boldsymbol{n}_{c,ij} \\ \boldsymbol{b}_{d,ij} + \boldsymbol{n}_{d,ij} \end{bmatrix} \tag{17}$$

where $\hat{\boldsymbol{a}}_{c,ij}$ and $\hat{\boldsymbol{a}}_{d,ij}$ are called the calibrated CM and DM accelerations. The $\boldsymbol{A}_d$ matrix can now be formed using $\hat{\boldsymbol{a}}_{c,ij}$ and $\hat{\boldsymbol{a}}_{d,ij}$ according to Eq. (8) and the error consists of a $\Delta\hat{\boldsymbol{M}}_{ij}^{-1}$ related term and linear combination of the accelerometer biases and noises.

In the final step of GGT retrieval, the accurate knowledge of centrifugal accelerations is required, as seen in Eq. (12). The angular velocities have been derived from an optimized combination of angular accelerations from the gradiometer, as seen in Eq. (10), and attitude quaternions from the star trackers [35,36]. The angular velocity estimation error will definitely affect the accuracy of the GGT measurements.

Based on the analysis of error sources, the GGT observation error can be decomposed as

$$\Delta\boldsymbol{V} = \boldsymbol{N}_{ICM} + \boldsymbol{B}_a + \boldsymbol{N}_a + \boldsymbol{N}_\omega \tag{18}$$

with

$$\boldsymbol{N}_{ICM} = -\left( \begin{bmatrix} \left(\Delta\hat{\boldsymbol{M}}_{c,14}^{-1}\boldsymbol{a}_{c,14} + \Delta\hat{\boldsymbol{M}}_{d,14}^{-1}\boldsymbol{a}_{d,14}\right)^T \\ \left(\Delta\hat{\boldsymbol{M}}_{c,25}^{-1}\boldsymbol{a}_{c,25} + \Delta\hat{\boldsymbol{M}}_{d,25}^{-1}\boldsymbol{a}_{d,25}\right)^T \\ \left(\Delta\hat{\boldsymbol{M}}_{c,36}^{-1}\boldsymbol{a}_{c,36} + \Delta\hat{\boldsymbol{M}}_{d,36}^{-1}\boldsymbol{a}_{d,36}\right)^T \end{bmatrix} + \begin{bmatrix} \left(\Delta\hat{\boldsymbol{M}}_{c,14}^{-1}\boldsymbol{a}_{c,14} + \Delta\hat{\boldsymbol{M}}_{d,14}^{-1}\boldsymbol{a}_{d,14}\right)^T \\ \left(\Delta\hat{\boldsymbol{M}}_{c,25}^{-1}\boldsymbol{a}_{c,25} + \Delta\hat{\boldsymbol{M}}_{d,25}^{-1}\boldsymbol{a}_{d,25}\right)^T \\ \left(\Delta\hat{\boldsymbol{M}}_{c,36}^{-1}\boldsymbol{a}_{c,36} + \Delta\hat{\boldsymbol{M}}_{d,36}^{-1}\boldsymbol{a}_{d,36}\right)^T \end{bmatrix}^T \right) \boldsymbol{L}^{-1} \tag{19}$$

$$\boldsymbol{B}_a = -\left( \begin{bmatrix} \left(\hat{\boldsymbol{M}}_{c,14}^{-1}\boldsymbol{b}_{c,14} + \hat{\boldsymbol{M}}_{d,14}^{-1}\boldsymbol{b}_{d,14}\right)^T \\ \left(\hat{\boldsymbol{M}}_{c,25}^{-1}\boldsymbol{b}_{c,25} + \hat{\boldsymbol{M}}_{d,25}^{-1}\boldsymbol{b}_{d,25}\right)^T \\ \left(\hat{\boldsymbol{M}}_{c,36}^{-1}\boldsymbol{b}_{c,36} + \hat{\boldsymbol{M}}_{d,36}^{-1}\boldsymbol{b}_{d,36}\right)^T \end{bmatrix} + \begin{bmatrix} \left(\hat{\boldsymbol{M}}_{c,14}^{-1}\boldsymbol{b}_{c,14} + \hat{\boldsymbol{M}}_{d,14}^{-1}\boldsymbol{b}_{d,14}\right)^T \\ \left(\hat{\boldsymbol{M}}_{c,25}^{-1}\boldsymbol{b}_{c,25} + \hat{\boldsymbol{M}}_{d,25}^{-1}\boldsymbol{b}_{d,25}\right)^T \\ \left(\hat{\boldsymbol{M}}_{c,36}^{-1}\boldsymbol{b}_{c,36} + \hat{\boldsymbol{M}}_{d,36}^{-1}\boldsymbol{b}_{d,36}\right)^T \end{bmatrix}^T \right) \boldsymbol{L}^{-1} \tag{20}$$

$$\boldsymbol{N}_a = -\left( \begin{bmatrix} \left(\hat{\boldsymbol{M}}_{c,14}^{-1}\boldsymbol{n}_{c,14} + \hat{\boldsymbol{M}}_{d,14}^{-1}\boldsymbol{n}_{d,14}\right)^T \\ \left(\hat{\boldsymbol{M}}_{c,25}^{-1}\boldsymbol{n}_{c,25} + \hat{\boldsymbol{M}}_{d,25}^{-1}\boldsymbol{n}_{d,25}\right)^T \\ \left(\hat{\boldsymbol{M}}_{c,36}^{-1}\boldsymbol{n}_{c,36} + \hat{\boldsymbol{M}}_{d,36}^{-1}\boldsymbol{n}_{d,36}\right)^T \end{bmatrix} + \begin{bmatrix} \left(\hat{\boldsymbol{M}}_{c,14}^{-1}\boldsymbol{n}_{c,14} + \hat{\boldsymbol{M}}_{d,14}^{-1}\boldsymbol{n}_{d,14}\right)^T \\ \left(\hat{\boldsymbol{M}}_{c,25}^{-1}\boldsymbol{n}_{c,25} + \hat{\boldsymbol{M}}_{d,25}^{-1}\boldsymbol{n}_{d,25}\right)^T \\ \left(\hat{\boldsymbol{M}}_{c,36}^{-1}\boldsymbol{n}_{c,36} + \hat{\boldsymbol{M}}_{d,36}^{-1}\boldsymbol{n}_{d,36}\right)^T \end{bmatrix}^T \right) \boldsymbol{L}^{-1} \tag{21}$$

$$\boldsymbol{N}_\omega = \begin{bmatrix} -2\left(\omega_y\Delta\omega_y + \omega_z\Delta\omega_z\right) & \omega_x\Delta\omega_y + \omega_y\Delta\omega_x & \omega_x\Delta\omega_z + \omega_z\Delta\omega_x \\ \omega_x\Delta\omega_y + \omega_y\Delta\omega_x & -2\left(\omega_x\Delta\omega_x + \omega_z\Delta\omega_z\right) & \omega_y\Delta\omega_z + \omega_z\Delta\omega_y \\ \omega_x\Delta\omega_z + \omega_z\Delta\omega_x & \omega_y\Delta\omega_z + \omega_z\Delta\omega_y & -2\left(\omega_x\Delta\omega_x + \omega_y\Delta\omega_y\right) \end{bmatrix} \tag{22}$$

where $\boldsymbol{N}_{ICM}$ is the noise due to the ICM calibration error, $\boldsymbol{B}_a$ and $\boldsymbol{N}_a$ are the bias and noise due to biases and noises in the six accelerometers, respectively, $\boldsymbol{N}_\omega$ is the noise due to angular velocity estimation error, $\Delta\hat{\boldsymbol{M}}_{c,ij}^{-1}$ and $\Delta\hat{\boldsymbol{M}}_{d,ij}^{-1}$ are the upper and lower 3×3 submatrices of $\Delta\hat{\boldsymbol{M}}_{ij}^{-1}$, $\hat{\boldsymbol{M}}_{c,ij}^{-1}$, and $\hat{\boldsymbol{M}}_{d,ij}^{-1}$ are the upper and lower 3×3 submatrices of $\hat{\boldsymbol{M}}_{ij}^{-1}$, and $\Delta\omega_x$, $\Delta\omega_y$, and $\Delta\omega_z$ are the angular velocity errors. The characteristics of these errors will be discussed in the next section.

## 3. Orbit determination models

This section describes the orbital dynamic model and the gravity gradient observation model used for GOCE orbit determination. In addition, the characteristics of the sources of measurement error are analyzed and a simplified measurement error model is formulated.

### *3.1. Orbital dynamic model*

The spacecraft's orbital motion is described by the following first-order differential equation

$$\frac{d}{dt}\begin{bmatrix} \boldsymbol{r} \\ \boldsymbol{v} \end{bmatrix} = \begin{bmatrix} \boldsymbol{v} \\ \boldsymbol{a}(t, \boldsymbol{r}, \boldsymbol{v}) \end{bmatrix} \tag{23}$$

where $\boldsymbol{r}$ and $\boldsymbol{v}$ are the inertial position and velocity vectors and $\boldsymbol{a}(t, \boldsymbol{r}, \boldsymbol{v})$ is the inertial acceleration. In this study, the International Celestial Reference Frame (ICRF) is used as the inertial coordinate system. For the definition of ICRF we refer to Petit et al. [37]. Given initial values and accurate force models, position and velocity over time can be obtained by numerical integration of Eq. (23).

The forces on LEO satellites usually include the Earth's gravitational attraction, third-body attractions from the Sun and the Moon, atmospheric drag, solar radiation pressure and thruster forces. In the case of GOCE, the non-gravitational forces in the flight direction were continuously compensated by electric propulsion. The remaining perturbation acceleration is on the order of $10^{-7}$ m/s$^2$. In this study, a 70×70 subset of the EGM2008 model is used to compute the acceleration due to the Earth's static gravity field. The tidal effects such as solid Earth tides, polar tides and ocean tides are not considered. The gravitational attractions of the Sun and Moon are modeled by using analytical series expansions of luni-solar coordinates [38].

The linearization of orbital equation requires the state transmission matrix, which refers to the partial derivative of the orbital state (position and velocity) at arbitrary time $t$ with respect to the initial state. The state transmission matrix is obtained by integration of the following differential equation

$$\frac{d}{dt}\boldsymbol{\Phi}(t,t_0) = \begin{bmatrix} \boldsymbol{0}_{3\times3} & \boldsymbol{I}_{3\times3} \\ \dfrac{\partial \boldsymbol{a}(t,\boldsymbol{r},\boldsymbol{v})}{\partial \boldsymbol{r}} & \dfrac{\partial \boldsymbol{a}(t,\boldsymbol{r},\boldsymbol{v})}{\partial \boldsymbol{v}} \end{bmatrix} \boldsymbol{\Phi}(t,t_0) \tag{24}$$

where $\boldsymbol{\Phi}(t,t_0)$ denotes the state transmission matrix from $t_0$ to $t$. The initial value of the equation above is

$$\boldsymbol{\Phi}(t_0,t_0) = \boldsymbol{I}_{6\times6} \tag{25}$$

The accuracy requirement for $\boldsymbol{\Phi}(t,t_0)$ is not as stringent as that for the trajectory integration. A simplification of the force model is used, where only the contribution of the Earth's gravitation up to degree 2 and order 0 is considered.

### *3.2. Gravity gradient observation model*

The gravity gradients are second-order derivatives of the gravitational potential with respect to position and form a 3×3 tensor matrix. The gravity gradients measured by GOCE are expressed in the GRF frame, whereas most of the Earth gravity models utilize the Earth-Centered Earth-Fixed (ECEF) frame. The GGT in GRF and the GGT in ECEF have the following relationship

$$\boldsymbol{V} = \boldsymbol{C} \cdot \boldsymbol{\Gamma} \cdot \boldsymbol{C}^T \tag{26}$$

with

$$\boldsymbol{V} = \begin{bmatrix} V_{xx} & V_{xy} & V_{xz} \\ V_{xy} & V_{yy} & V_{yz} \\ V_{xz} & V_{yz} & V_{zz} \end{bmatrix}, \quad \boldsymbol{\Gamma} = \begin{bmatrix} \Gamma_{xx} & \Gamma_{xy} & \Gamma_{xz} \\ \Gamma_{xy} & \Gamma_{yy} & \Gamma_{yz} \\ \Gamma_{xz} & \Gamma_{yz} & \Gamma_{zz} \end{bmatrix} \tag{27}$$

where $\boldsymbol{V}$ is the GGT in GRF, $\boldsymbol{\Gamma}$ is the GGT in ECEF, and $\boldsymbol{C}$ is the rotation matrix from ECEF to GRF. The coordinate transformation is implemented via the ECI frame. The star trackers provide accurate attitude information which can be used to compute the rotation matrix from ECI to GRF. In this study, we use the ICRF/ITRF2008 coordinate frame for the definition of ECI and ECEF and the IERS models provide high-precision coordinate transformation [37].

The GGT is a symmetric matrix and contains 9 elements. As mentioned in Section 2, the symmetric property of GGT is used to isolate Euler accelerations. Thus the GOCE gradiometer outputs only 6 components at each epoch. The measurement equation is usually given in a vector form. Rewrite $\boldsymbol{V}$ and $\boldsymbol{\Gamma}$ into column vectors

$$\vec{\boldsymbol{V}} = \begin{bmatrix} V_{xx} \\ V_{yy} \\ V_{zz} \\ V_{xy} \\ V_{xz} \\ V_{yz} \end{bmatrix}, \quad \vec{\boldsymbol{\Gamma}} = \begin{bmatrix} \Gamma_{xx} \\ \Gamma_{yy} \\ \Gamma_{zz} \\ \Gamma_{xy} \\ \Gamma_{xz} \\ \Gamma_{yz} \end{bmatrix} \tag{28}$$

where the array symbol denotes the vector form of GGT. The relationship between $\vec{\boldsymbol{V}}$ and $\vec{\boldsymbol{\Gamma}}$ is

$$\vec{\boldsymbol{V}} = \boldsymbol{\Pi} \cdot \vec{\boldsymbol{\Gamma}} \tag{29}$$

with

$$\boldsymbol{\Pi} = \begin{bmatrix} c_{11}^2 & c_{12}^2 & c_{13}^2 & 2c_{11}c_{12} & 2c_{11}c_{13} & 2c_{12}c_{13} \\ c_{21}^2 & c_{22}^2 & c_{23}^2 & 2c_{21}c_{22} & 2c_{21}c_{23} & 2c_{22}c_{23} \\ c_{31}^2 & c_{32}^2 & c_{33}^2 & 2c_{31}c_{32} & 2c_{31}c_{33} & 2c_{32}c_{33} \\ c_{11}c_{21} & c_{12}c_{22} & c_{13}c_{23} & c_{12}c_{21}+c_{11}c_{22} & c_{13}c_{21}+c_{11}c_{23} & c_{13}c_{22}+c_{12}c_{23} \\ c_{11}c_{31} & c_{12}c_{32} & c_{13}c_{33} & c_{12}c_{31}+c_{11}c_{32} & c_{13}c_{31}+c_{11}c_{33} & c_{13}c_{32}+c_{12}c_{33} \\ c_{21}c_{31} & c_{22}c_{32} & c_{23}c_{33} & c_{22}c_{31}+c_{21}c_{32} & c_{23}c_{31}+c_{21}c_{33} & c_{23}c_{32}+c_{22}c_{33} \end{bmatrix} \tag{30}$$

where the coefficient matrix $\boldsymbol{\Pi}$ comprises elements from the rotation matrix $\boldsymbol{C}$ and $c_{ij}$ represents the $i$th row and $j$th column element in $\boldsymbol{C}$.

The gravitational potential is expressed as a series of spherical harmonics [39]

$$U = \frac{GM}{R}\sum_{n=0}^{\infty}\sum_{m=0}^{n}\left(\bar{C}_{nm}\bar{V}_{nm} + \bar{S}_{nm}\bar{W}_{nm}\right) \tag{31}$$

with

$$\bar{V}_{nm} = \left(\frac{R}{r}\right)^{n+1} \bar{P}_{nm}\left(\sin\phi\right)\cos m\lambda \tag{32}$$

$$\bar{W}_{nm} = \left(\frac{R}{r}\right)^{n+1} \bar{P}_{nm}\left(\sin\phi\right)\sin m\lambda \tag{33}$$

where $GM$ is the Earth's geocentric gravitational constant, $R$ is the reference equatorial radius of the Earth, $n$ and $m$ are degree and order, $\bar{C}_{nm}$ and $\bar{S}_{nm}$ are normalized spherical harmonic coefficients, $\bar{V}_{nm}$ and $\bar{W}_{nm}$ are the associated normalized terms, $\bar{P}_{nm}$ is the normalized associated Legendre function of the first kind, and $r$, $\phi$, and $\lambda$ are the geocentric distance, latitude, and longitude of position in ECEF. $GM$, $R$, $\bar{C}_{nm}$, and $\bar{S}_{nm}$ are constants and their values are provided in Earth gravity model files. The recursive computation of $\bar{V}_{nm}$ and $\bar{W}_{nm}$ are given in Montenbruck et al. [38].

The expression of the GGT in ECEF can be obtained by evaluating the second-order derivatives of $U$

$$\Gamma_{ij} = \frac{GM}{R}\sum_{n=0}^{\infty}\sum_{m=0}^{n}\left(\bar{C}_{nm}\frac{\partial^2\bar{V}_{nm}}{\partial i\partial j} + \bar{S}_{nm}\frac{\partial^2\bar{W}_{nm}}{\partial i\partial j}\right) \tag{34}$$

where $ij \in \{xx, yy, zz, xy, xz, yz\}$ represents the index of the gravity gradient components. The unit of GGT is Eötvös, denoted by the symbol E. 1 E equals $10^{-9}$ $1/s^2$ in SI units. In this study, a 120×120 subset of the EGM2008 gravity model is used to compute gravity gradients. The contributions of tidal effects are on the order of 0.1 mE (milli-Eötvös) and are thus not considered [40].

The partial derivatives of the GGT in ECEF with respect to position are components of the third-order gravity tensor and can be calculated by

$$T_{ij,k} = \frac{GM}{R}\sum_{n=0}^{\infty}\sum_{m=0}^{n}\left(\bar{C}_{nm}\frac{\partial^3\bar{V}_{nm}}{\partial i\partial j\partial k} + \bar{S}_{nm}\frac{\partial^3\bar{W}_{nm}}{\partial i\partial j\partial k}\right) \tag{35}$$

where $k \in \{x, y, z\}$ represents the index of the position components. The properties of the third-order gravity tensor could be found in Šprlák and Novák [41]. The third-order gravity tensor contains 27 components, which can be used to compute the gravity gradient Jacobian matrix, i.e., the partial derivative matrix of $\vec{\boldsymbol{\Gamma}}$ with respect to position. The accuracy requirement for the computation of $T_{ij,k}$ is also not stringent and only the Earth's gravitation up to degree 2 and order 0 is involved.

### *3.3. Measurement error model*

As stated in Section 2, the errors of GOCE gravity gradient measurements are composed of several parts. An appropriate error model is essential to the formulation of the measurement equation. The following is an analysis of the characteristics of the errors in Eq. (18), from perspectives of frequency spectrum and noise level.

First consider the ICM calibration-induced noise $\boldsymbol{N}_{ICM}$. According to the error budget analysis given in Cesare et al. [34], the submatrix $\Delta\hat{\boldsymbol{M}}_{c,ij}^{-1}$ is close enough to zero and the elements in $\Delta\hat{\boldsymbol{M}}_{d,ij}^{-1}$ have a maximum value of about $3\times10^{-3}$. GOCE data show that the differential-mode accelerations $\boldsymbol{a}_{d,ij}$ are nearly constant and are disturbed by small periodical variations. The variations are concentrated near the orbit revolution frequency and their magnitudes are between 10 and 100 E. Therefore, the noise $\boldsymbol{N}_{ICM}$ can be modeled as a constant bias plus small periodical variations (between 0.03 and 0.3 E).

The bias $\boldsymbol{B}_a$ and the noise $\boldsymbol{N}_a$ are determined by the characteristic of the accelerometers. According to Rummel et al. [8], the accelerometers achieve high sensitivity only in frequencies between $5\times10^{-3}$ and 0.1 Hz. Inside the measurement bandwidth, the accelerometers show white noise behavior. Along the ultra-sensitive axes, the noise density level is about $10^{-12}$ m/s$^2$/√Hz, whereas along the less sensitive axes, the noise density level is about $3\times10^{-10}$ m/s$^2$/√Hz [33]. Below the measurement bandwidth, the noise is proportional to the inverse of the frequency, and shows a typical drift behavior in the time domain. As seen in Eqs. (20) and (21), $\boldsymbol{B}_a$ and $\boldsymbol{N}_a$ are linear functions of the accelerometer biases and noises. Therefore, the sum of the two errors can be modeled as a drifting bias plus white noise.

The characteristic of the noise $\boldsymbol{N}_\omega$ is determined by the angular velocities and their estimation errors. Similar to the differential-mode accelerations, the angular velocities of GOCE are also nearly constant and contain periodical variations having a magnitude of about $10^{-4}$ rad/s. Due to the spectral combination of the angular accelerations and the attitude quaternions, the angular velocity errors show a $f$ behavior in lower frequencies and show a $1/f^2$ behavior in higher frequencies. The maximum value of the angular velocity errors is about $10^{-6}$ rad/s [33]. Therefore, according to Eq. (22), the periodical variations of $\boldsymbol{N}_{\omega}$ are on the order of 0.1 E ($10^{-10}$ $s^{-2}$).

By summing up all the errors above, the noise characteristic of the total observation error $\Delta \boldsymbol{V}$ can be obtained. Inside the measurement bandwidth, the noise shows white noise behavior. For the $V_{xx}, V_{yy}, V_{zz}$ and $V_{xz}$ components, the noise density levels are on the order of 10 mE√Hz, whereas for the $V_{xy}$ and $V_{yz}$ components, the noise density levels are much higher, 350 mE√Hz and 500 mE√Hz, respectively. Below the measurement bandwidth, the noise increases inversely with frequency and is superimposed by cyclic distortions (due to the variations of $\boldsymbol{N}_{ICM}$ and $\boldsymbol{N}_{\omega}$). Fig. 2 plots the measurement error of the $V_{yy}$ component as an example.

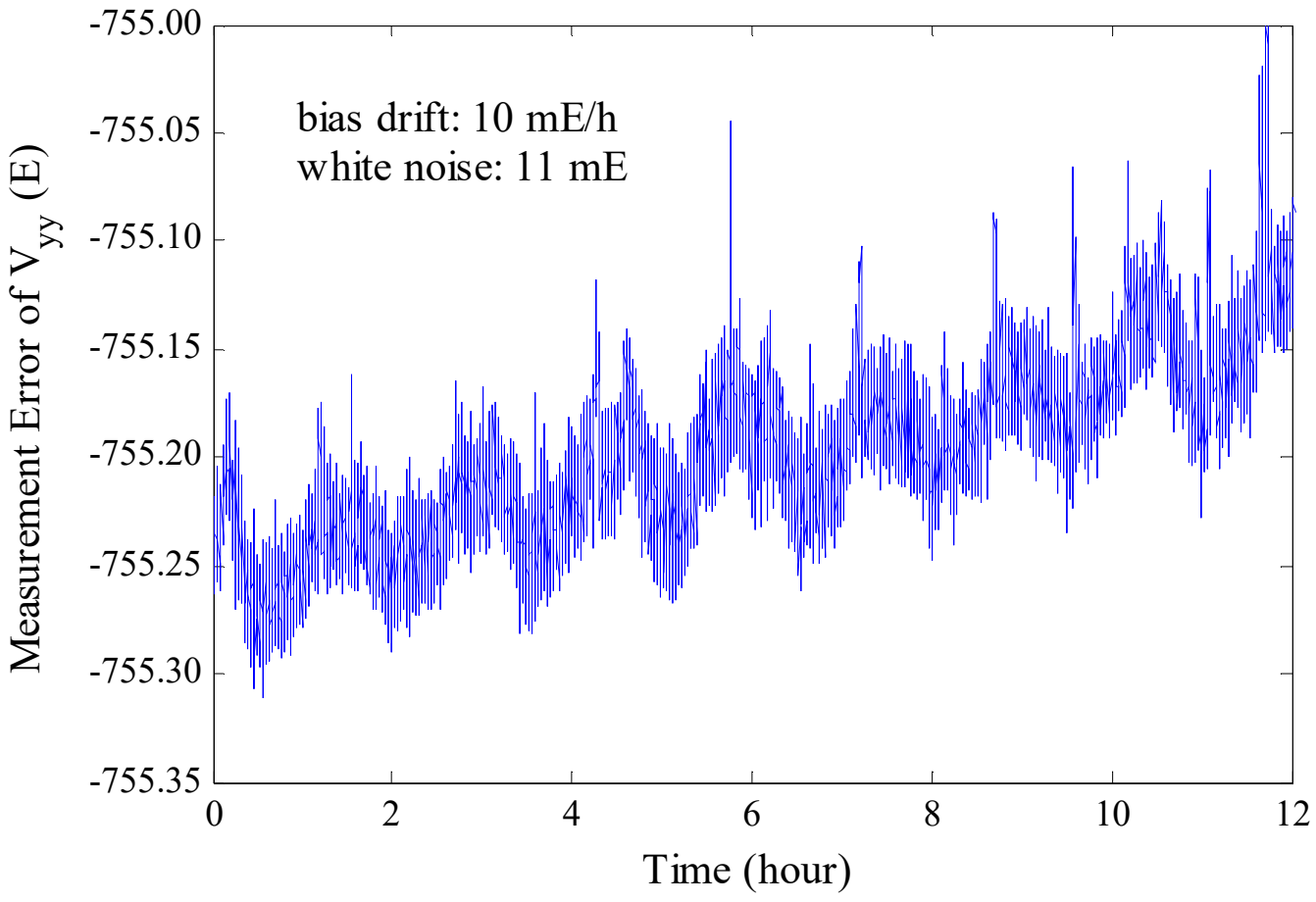

**Fig. 2**. GOCE Gravity gradient measurement error varying with time; $V_{yy}$ component from 8 September 2013.

Therefore, the observation error of each gravity gradient component in this study is modeled as a drifting bias plus small periodical variations as well as low-level white noise

$$\Delta V_{ij}(t) = b_{ij} + d_{ij}(t - t_0) + n_{ij} + v_{ij} \tag{36}$$

where $ij$ denotes the index of the gravity gradient component, $b_{ij}$ is the bias at the reference epoch $t_0$, $d_{ij}$ is the constant drift of the bias, $n_{ij}$ is the low-frequency noise, and $v_{ij}$ is the white noise. The bias $b_{ij}$ represents the sum of $\boldsymbol{B}_a$ and the constant parts of $\boldsymbol{N}_{ICM}$ and $\boldsymbol{N}_{\omega}$. The drift $d_{ij}$ is due to the $1/f$ behavior of the accelerometer noises.

The noise $n_{ij}$ represents the sum of periodical variations of $\boldsymbol{N}_{ICM}$ and $\boldsymbol{N}_{\omega}$ and has a magnitude of about 0.1 E and $v_{ij}$ represents the white noise part of $\boldsymbol{N}_a$ .

Let the vector $\boldsymbol{z}$ denote the gravity gradient measurements in the GRF frame. The observation equation can be written as

$$\boldsymbol{z}(t) = \vec{\boldsymbol{V}} + \Delta\vec{\boldsymbol{V}} = \boldsymbol{\Pi} \cdot \vec{\boldsymbol{\Gamma}} + \boldsymbol{b} + \boldsymbol{d}(t - t_0) + \boldsymbol{n} + \boldsymbol{v} \tag{37}$$

where $\boldsymbol{b}$ is the bias vector, $\boldsymbol{d}$ is the constant drift vector, $\boldsymbol{n}$ is the low-frequency noise, and $\boldsymbol{v}$ is the white noise.

## 4. Batch filter design

The orbital motion is close to linear over a small range, whereas the gravity gradient observations are highly nonlinear functions. A hybrid least squares (HLS) batch filter based on linearization of the orbital equation and unscented transformation of the measurement equation is developed to deal with different degrees of nonlinearity in the system. In addition, to restrict orbit linearization errors, the filter adaptively adjusts the time span of the measurement data arc at each iteration step. This section starts with an overview of the iterated least squares and the unscented least squares, and a detailed description of the adaptive HLS filter design is given thereafter.

### *4.1. The iterated least squares and the unscented least squares*

The principle of least squares (LS) batch filter is to determine a set of states that minimizes the sum of the squares of measurement residuals. For nonlinear measurement equations, the iterated least squares (ILS) filter iteratively improves state estimation using the first-order partial derivatives of the linearized system [42]. For highly nonlinear functions, sigma-point transformation is introduced in Park et al. [43], and a non-recursive unscented least squares (ULS) filter is developed. To illustrate the methods of ILS and ULS, the measurement equation is redefined in the following general mathematical form

$$\boldsymbol{z} = \begin{bmatrix} z_1 \\ z_2 \\ \vdots \\ z_n \end{bmatrix} = \begin{bmatrix} h_1(\boldsymbol{x}) \\ h_2(\boldsymbol{x}) \\ \vdots \\ h_n(\boldsymbol{x}) \end{bmatrix} + \begin{bmatrix} w_1 \\ w_2 \\ \vdots \\ w_n \end{bmatrix} = \boldsymbol{h}(\boldsymbol{x}) + \boldsymbol{w} \tag{38}$$

where $\boldsymbol{x}$ is the state vector, $\boldsymbol{z}$ is the stacked vector of measurements, $\boldsymbol{h}$ is the stacked measurement function, and $\boldsymbol{w}$ is the additive zero-mean measurement noise vector.

Assume the initial state estimate and its covariance as

$$\hat{\boldsymbol{x}}_j^{\mathrm{ILS}} = \hat{\boldsymbol{x}}_0,\ \hat{\boldsymbol{P}}_{\boldsymbol{xx},j}^{\mathrm{ILS}} = \hat{\boldsymbol{P}}_0,\ j = 0 \tag{39}$$

where $\hat{\boldsymbol{x}}_0$ is supposed to be a priori information about the state, $\hat{\boldsymbol{P}}_0$ is a covariance matrix representing the uncertainty of $\hat{\boldsymbol{x}}_0$, and $j$ is the iteration number. It should be noted that the subscript $\boldsymbol{xx}$ implies the covariance matrix of the state vector $\boldsymbol{x}$.

The ILS filter updates the estimate $\hat{\boldsymbol{x}}_j^{\mathrm{ILS}}$ as follows

$$\hat{\boldsymbol{x}}_{j+1}^{\mathrm{ILS}} = \hat{\boldsymbol{x}}_j^{\mathrm{ILS}} + \left(\hat{\boldsymbol{P}}_0^{-1} + \boldsymbol{H}_j^T \boldsymbol{R}^{-1} \boldsymbol{H}_j\right)^{-1} \left[\hat{\boldsymbol{P}}_0^{-1}\left(\hat{\boldsymbol{x}}_j^{\mathrm{ILS}} - \hat{\boldsymbol{x}}_0\right) + \boldsymbol{H}_j^T \boldsymbol{R}^{-1}\left[\boldsymbol{z} - \boldsymbol{h}\left(\hat{\boldsymbol{x}}_j^{\mathrm{ILS}}\right)\right]\right] \tag{40}$$

where

$$\boldsymbol{H}_j = \left.\frac{\partial \boldsymbol{h}(\boldsymbol{x})}{\partial \boldsymbol{x}}\right|_{\boldsymbol{x}=\hat{\boldsymbol{x}}_j^{\mathrm{ILS}}} \tag{41}$$

is the measurement Jacobian matrix and $\boldsymbol{R}$ is the covariance matrix of the noise $\boldsymbol{w}$. The convergence criterion for the iteration is usually given by

$$\frac{\left|\hat{\boldsymbol{x}}_{j+1}^{\mathrm{ILS}} - \hat{\boldsymbol{x}}_j^{\mathrm{ILS}}\right|}{\left|\hat{\boldsymbol{x}}_j^{\mathrm{ILS}}\right|} \le \eta \tag{42}$$

where $\eta$ is a predefined relative error tolerance. The covariance of the ILS final state estimate is

$$\hat{\boldsymbol{P}}_{\boldsymbol{xx}}^{\mathrm{ILS}} = \left(\hat{\boldsymbol{P}}_0^{-1} + \boldsymbol{H}_j^T \boldsymbol{R}^{-1} \boldsymbol{H}_j\right)^{-1} \tag{43}$$

The ILS filter performs well for weakly nonlinear equations. For highly nonlinear systems, however, the linear approximation will induce significant errors and make it difficult to achieve convergence. The ULS filter is an extension of the sequential unscented Kalman filter and deals with the nonlinearity problem using a set of selected sigma points. The mean and the covariance of the measurements are calculated and are used to correct the state estimate.

Assume the initial state estimate and its covariance as

$$\hat{\boldsymbol{x}}_j^{\mathrm{ULS}} = \hat{\boldsymbol{x}}_0,\ \hat{\boldsymbol{P}}_{\boldsymbol{xx},j}^{\mathrm{ULS}} = \hat{\boldsymbol{P}}_0,\ j = 0 \tag{44}$$

where $j$ is the iteration number, and the subscript $\boldsymbol{xx}$ implies the covariance matrix of the state vector $\boldsymbol{x}$. The sigma points are selected as follows

$$
\begin{aligned}
\boldsymbol{\chi}_{0,j} &= \hat{\boldsymbol{x}}_j^{\mathrm{ULS}} \\
\boldsymbol{\chi}_{i,j} &= \hat{\boldsymbol{x}}_j^{\mathrm{ULS}} + \left(\sqrt{\left(L+\zeta\right)\hat{\boldsymbol{P}}_{\boldsymbol{xx},j}^{\mathrm{ULS}}}\right)_i, \quad i = 1,\ldots,L \\
\boldsymbol{\chi}_{i,j} &= \hat{\boldsymbol{x}}_j^{\mathrm{ULS}} - \left(\sqrt{\left(L+\zeta\right)\hat{\boldsymbol{P}}_{\boldsymbol{xx},j}^{\mathrm{ULS}}}\right)_i, \quad i = L+1,\ldots,2L
\end{aligned}
\tag{45}
$$

where $L$ is the dimension of $\boldsymbol{x}$, $\zeta = \alpha^2\left(L+\kappa\right) - L$ is a scaling parameter, $\alpha$ is a constant and is usually set to a small positive value, $\kappa$ is a secondary scaling parameter which is usually set to 0 or 3-$L$, and $\left(\sqrt{\left(L+\zeta\right)\hat{\boldsymbol{P}}_{\boldsymbol{xx},j}^{\mathrm{ULS}}}\right)_i$ is the $i$th column of the square root of $\left(L+\zeta\right)\hat{\boldsymbol{P}}_{\boldsymbol{xx},j}^{\mathrm{ULS}}$.

Each sigma point is propagated using the nonlinear measurement function

$$
\boldsymbol{\gamma}_{i,j} = \boldsymbol{h}\left(\boldsymbol{\chi}_{i,j}\right), \quad i = 0,\ldots,2L \tag{46}
$$

The mean and the covariance of the measurement vector are calculated as follows

$$
\bar{\boldsymbol{z}}_j = \sum_{i=0}^{2L} W_i^{(m)} \boldsymbol{\gamma}_{i,j} \tag{47}
$$

$$
\hat{\boldsymbol{P}}_{\boldsymbol{zz},j} = \sum_{i=0}^{2L} W_i^{(c)} \left(\boldsymbol{\gamma}_{i,j} - \bar{\boldsymbol{z}}_j\right)\left(\boldsymbol{\gamma}_{i,j} - \bar{\boldsymbol{z}}_j\right)^T + \boldsymbol{R} \tag{48}
$$

where the subscript $\boldsymbol{zz}$ implies the covariance matrix of the measurement vector $\boldsymbol{z}$. The cross-correlation matrix of $\boldsymbol{x}$ and $\boldsymbol{z}$ is

$$
\hat{\boldsymbol{P}}_{\boldsymbol{xz},j} = \sum_{i=0}^{2L} W_i^{(c)} \left(\boldsymbol{\chi}_{i,j} - \hat{\boldsymbol{x}}_j^{\mathrm{ULS}}\right)\left(\boldsymbol{\gamma}_{i,j} - \bar{\boldsymbol{z}}_j\right)^T \tag{49}
$$

Where the subscript $\boldsymbol{xz}$ implies the cross-covariance matrix of the state vector $\boldsymbol{x}$ and the measurement vector $\boldsymbol{z}$. $W_i^{(m)}$ and $W_i^{(c)}$ are the weighting factors and are defined as follows

$$
\begin{aligned}
W_0^{(m)} &= \frac{\zeta}{L+\zeta} \\
W_0^{(c)} &= \frac{\zeta}{L+\zeta} + \left(1 - \alpha^2 + \beta\right) \\
W_i^{(m)} &= W_i^{(c)} = \frac{1}{2\left(L+\zeta\right)}, \quad i = 1,\ldots,2L
\end{aligned}
\tag{50}
$$

where $\beta$ is the third scaling parameter and is used to incorporate prior knowledge of the distribution of $\boldsymbol{x}$.

The ULS filter updates the estimate $\hat{\boldsymbol{x}}_j^{\mathrm{ULS}}$ as follows

$$
\hat{\boldsymbol{x}}_{j+1}^{\mathrm{ULS}} = \hat{\boldsymbol{x}}_j^{\mathrm{ULS}} + \boldsymbol{K}_j\left(\boldsymbol{z} - \bar{\boldsymbol{z}}_j\right) \tag{51}
$$

with

$$\boldsymbol{K}_j = \hat{\boldsymbol{P}}_{xz,j}\hat{\boldsymbol{P}}_{zz,j}^{-1} \tag{52}$$

where $\boldsymbol{K}_j$ is the optimal filter gain.

The convergence criterion can be set to be the same as that of the ILS. The covariance of the final estimate is given by

$$\hat{\boldsymbol{P}}_{xx}^{\text{ULS}} = \hat{\boldsymbol{P}}_0 - \boldsymbol{K}_1\hat{\boldsymbol{P}}_{zz,1}\boldsymbol{K}_1^T \tag{53}$$

The covariance matrices $\hat{\boldsymbol{P}}_{xx}^{\text{ILS}}$ and $\hat{\boldsymbol{P}}_{xx}^{\text{ULS}}$ are consistent with the state estimation if and only if the measurement noise $\boldsymbol{w}$ is white and Gaussian. Otherwise, a fudge factor has to be added to guarantee the filter consistency

$$\hat{\boldsymbol{P}}_{xx}^{\text{ILS}} = \vartheta^{\text{ILS}}\left(\boldsymbol{H}_j^T\boldsymbol{R}^{-1}\boldsymbol{H}_j\right)^{-1} \tag{54}$$

$$\hat{\boldsymbol{P}}_{xx}^{\text{ULS}} = \vartheta^{\text{ULS}}\left(\hat{\boldsymbol{P}}_0 - \boldsymbol{K}_1\hat{\boldsymbol{P}}_{zz,1}\boldsymbol{K}_1^T\right) \tag{55}$$

The fudge factor should be set to a large value. Actually, the fudge factor affect the convergence process of the filter mainly.

*4.2. Adaptive hybrid least squares batch filter design*

The batch orbit determination problem is to estimate unknown orbital elements from a set of measurements. The measurement function $\boldsymbol{h}$ actually consists of not only the observation equation but also the orbital equation. The adaptive HLS filter exploits the different degrees of nonlinearity in the two equations and autonomously adjusts the measurement time span to bound orbit linearization errors. In this study, the adaptive HLS filter is first implemented to estimate the initial position and velocity as well as the initial biases. An additional augmented state ILS filter is then carried out to obtain the bias drifts and to correct the adaptive HLS filter's results. The low-frequency noise $\boldsymbol{n}$ is not estimated and is included with $\boldsymbol{v}$ into the measurement noise vector $\boldsymbol{w}$ as follows

$$\boldsymbol{w} = \begin{bmatrix} \boldsymbol{n}_0 + \boldsymbol{v}_0 \\ \boldsymbol{n}_1 + \boldsymbol{v}_1 \\ \vdots \\ \boldsymbol{n}_{N-1} + \boldsymbol{v}_{N-1} \end{bmatrix} \tag{56}$$

where $N$ is the total measurement epochs. The standard deviations of the six components of the noise ($\boldsymbol{n} + \boldsymbol{v}$) are set to be 100 mE, 100 mE, 100 mE, 350 mE, 100 mE, and 500 mE, respectively.

The algorithm of the adaptive HLS for GOCE gravity gradient based orbit determination is proposed as follows. The state vector $\hat{\boldsymbol{x}}^{\mathrm{HLS}}$ comprises the 6-dimensional orbital state vector $\boldsymbol{y}$ (the initial position $\boldsymbol{r}_0$ and the initial velocity $\boldsymbol{v}_0$) and the 6-dimensional bias vector $\boldsymbol{b}$. The initial state and its covariance are assumed as

$$\hat{\boldsymbol{x}}_j^{\mathrm{HLS}} = \hat{\boldsymbol{x}}_0,\ \hat{\boldsymbol{P}}_{\boldsymbol{xx},j}^{\mathrm{HLS}} = \hat{\boldsymbol{P}}_0,\ j = 0 \tag{57}$$

where $j$ is the iteration number. The orbital state $\boldsymbol{y}$ and the bias vector $\boldsymbol{b}$ and their covariances can be extracted as follows

$$\hat{\boldsymbol{y}}_j = \left[\hat{\boldsymbol{x}}_j^{\mathrm{HLS}}\right]_{1:6},\ \hat{\boldsymbol{P}}_{\boldsymbol{yy},j} = \left[\hat{\boldsymbol{P}}_{\boldsymbol{xx},j}^{\mathrm{HLS}}\right]_{1:6,1:6} \tag{58}$$

$$\hat{\boldsymbol{b}}_j = \left[\hat{\boldsymbol{x}}_j^{\mathrm{HLS}}\right]_{7:12},\ \hat{\boldsymbol{P}}_{\boldsymbol{bb},j} = \left[\hat{\boldsymbol{P}}_{\boldsymbol{xx},j}^{\mathrm{HLS}}\right]_{7:12,7:12} \tag{59}$$

where the indices refer to the elements of the vector and the matrix. The sigma points of the orbital state are selected by

$$\begin{aligned}
\chi_{0,j} &= \hat{\boldsymbol{y}}_j \\
\chi_{i,j} &= \hat{\boldsymbol{y}}_j + \left(\sqrt{(L+\zeta)\hat{\boldsymbol{P}}_{\boldsymbol{yy},j}}\right)_i,\ i = 1,\ldots,L \\
\chi_{i,j} &= \hat{\boldsymbol{y}}_j - \left(\sqrt{(L+\zeta)\hat{\boldsymbol{P}}_{\boldsymbol{yy},j}}\right)_i,\ i = L+1,\ldots,2L
\end{aligned} \tag{60}$$

In the case of GOCE orbit determination, $L = 6$.

The sigma points are propagated using the orbital integrator and the state transmission matrices to obtain the position at each measurement epoch $t_k\ (k = 0,\ldots,N_j - 1)$

$$\begin{aligned}
\psi_{0,k,j} &= \left[\boldsymbol{f}(\chi_{0,j}, t_k)\right]_{1:3} \\
\psi_{i,k,j} &= \psi_{0,k,j} + \left[\boldsymbol{\Phi}(t_k, t_0)\right]_{1:3,1:6}(\chi_{i,j} - \chi_{0,j}),\ i = 1,\ldots,2L
\end{aligned} \tag{61}$$

where $\psi_{k,i,j}\ (i = 0,\ldots,2L)$ are the sigma points of position at epoch $t_k$ and the nonlinear function $\boldsymbol{f}$ represents the orbital integrator. In this study, a variable-order Adams-Bashforth-Moulton integrator is used. The calculation of $\psi_{k,i,j}\ (i = 1,\ldots,2L)$ makes use of linearization of the orbital equation. To restrict the linearization errors, the time span $N_j$ is determined from the position uncertainties. The covariance of position at epoch $t_k$ is

$$\hat{\boldsymbol{P}}_{\psi\psi,k,j} = \left[\boldsymbol{\Phi}(t_k, t_0)\right]_{1:3,1:6} \hat{\boldsymbol{P}}_{\boldsymbol{yy},j} \left[\boldsymbol{\Phi}^T(t_k, t_0)\right]_{1:6,1:3} \tag{62}$$

The $N_j$ is determined by

$$N_j = \max_N \left( \left\| \hat{\boldsymbol{P}}_{\boldsymbol{\psi\psi},N,j} \right\|_p <= \varepsilon \right) \tag{63}$$

where $\varepsilon$ is the tolerance of the position uncertainty and the operator $\| \cdot \|_p$ is defined as

$$\left\| \boldsymbol{A}_{3\times3} \right\|_p = \sqrt{A_{11}^2 + A_{22}^2 + A_{33}^2} \tag{64}$$

The sigma points of positions are further propagated to obtain the stacked gravity gradient measurements as follows

$$\boldsymbol{\gamma}_{i,j} = \begin{bmatrix} \vec{\boldsymbol{V}}\left(\boldsymbol{\psi}_{i,0,j}\right) + \hat{\boldsymbol{b}}_j \\ \vec{\boldsymbol{V}}\left(\boldsymbol{\psi}_{i,1,j}\right) + \hat{\boldsymbol{b}}_j \\ \vdots \\ \vec{\boldsymbol{V}}\left(\boldsymbol{\psi}_{i,N_j,j}\right) + \hat{\boldsymbol{b}}_j \end{bmatrix}, \ i = 0,\ldots,2L \tag{65}$$

The mean and covariance of the gravity gradient measurements are

$$\overline{\boldsymbol{z}}_j = \sum_{i=0}^{2L} W_i^{(m)} \boldsymbol{\gamma}_{i,j} \tag{66}$$

$$\hat{\boldsymbol{P}}_{\boldsymbol{zz},j} = \sum_{i=0}^{2L} W_i^{(c)} \left(\boldsymbol{\gamma}_{i,j} - \overline{\boldsymbol{z}}_j\right)\left(\boldsymbol{\gamma}_{i,j} - \overline{\boldsymbol{z}}_j\right)^T + \boldsymbol{H}_{\boldsymbol{b}} \hat{\boldsymbol{P}}_{\boldsymbol{bb},j} \boldsymbol{H}_{\boldsymbol{b}}^T + \boldsymbol{R} \tag{67}$$

where $\boldsymbol{H}_{\boldsymbol{b}}$ is the partial derivative matrix of the stacked vector of measurements $\boldsymbol{z}$ with respect to the bias vector $\boldsymbol{b}$

$$\boldsymbol{H}_{\boldsymbol{b}} = \begin{bmatrix} \boldsymbol{I}_{6\times6} \\ \boldsymbol{I}_{6\times6} \\ \vdots \\ \boldsymbol{I}_{6\times6} \end{bmatrix}_{(6N_j)\times6} \tag{68}$$

and $\boldsymbol{R}$ is the covariance of the measurement noise vector $\boldsymbol{w}$.

The cross-correlation matrix of $\boldsymbol{y}$ and $\boldsymbol{z}$ is

$$\hat{\boldsymbol{P}}_{\boldsymbol{yz},j} = \sum_{i=0}^{2L} W_i^{(c)} \left(\boldsymbol{\chi}_{i,j} - \hat{\boldsymbol{y}}_j\right)\left(\boldsymbol{\gamma}_{i,j} - \overline{\boldsymbol{z}}_j\right)^T \tag{69}$$

and the cross-correlation matrix of $\boldsymbol{b}$ and $\boldsymbol{z}$ is

$$\hat{\boldsymbol{P}}_{\boldsymbol{bz},j} = \hat{\boldsymbol{P}}_{\boldsymbol{bb},j} \boldsymbol{H}_{\boldsymbol{b}}^T \tag{70}$$

Thus the cross-correlation matrix of $\boldsymbol{x}$ and $\boldsymbol{z}$ is

$$\hat{\boldsymbol{P}}_{\boldsymbol{xz},j} = \begin{bmatrix} \hat{\boldsymbol{P}}_{\boldsymbol{yz},j} \\ \hat{\boldsymbol{P}}_{\boldsymbol{bz},j} \end{bmatrix} \tag{71}$$

The HLS filter updates the estimate $\hat{\boldsymbol{x}}_j^{\text{HLS}}$ and its covariance $\hat{\boldsymbol{P}}_{\boldsymbol{xx},j}^{\text{HLS}}$ as follows

$$\hat{\boldsymbol{x}}_{j+1}^{\text{HLS}} = \hat{\boldsymbol{x}}_j^{\text{HLS}} + \boldsymbol{K}_j \left( \boldsymbol{z} - \overline{\boldsymbol{z}}_j \right) \tag{72}$$

$$\hat{\boldsymbol{P}}_{\boldsymbol{xx},j+1}^{\text{HLS}} = \vartheta^{\text{HLS}} \left( \hat{\boldsymbol{P}}_{\boldsymbol{xx},j}^{\text{HLS}} - \boldsymbol{K}_j \hat{\boldsymbol{P}}_{\boldsymbol{zz},j} \boldsymbol{K}_j^T \right) \tag{73}$$

with

$$\boldsymbol{K}_j = \hat{\boldsymbol{P}}_{\boldsymbol{xz},j} \hat{\boldsymbol{P}}_{\boldsymbol{zz},j}^{-1} \tag{74}$$

The iteration is terminated by the same convergence condition of the ILS filter in Eq. (42). The flowchart of the adaptive HLS filter is summarized in Fig. 3.

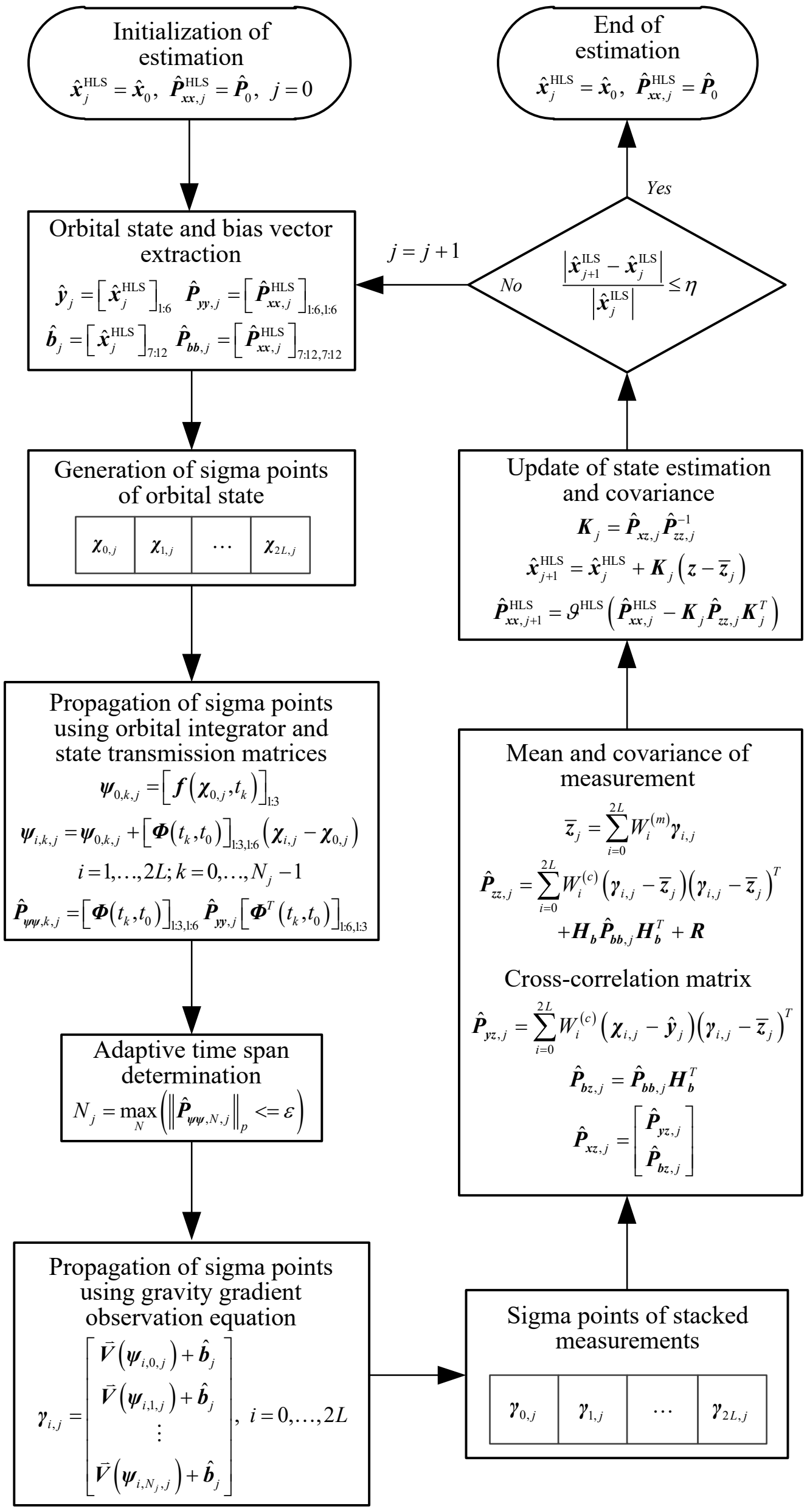

**Fig. 3**. Flowchart of the adaptive hybrid least squares.

After the adaptive HLS filter, an augmented state ILS filter is used to estimate the drifts and to improve the orbit determination results. The augmented state vector is defined as

$$\boldsymbol{x}^{a} = \begin{bmatrix} \boldsymbol{y} \\ \boldsymbol{b} \\ \boldsymbol{d} \end{bmatrix} \tag{75}$$

And the measurement Jacobian matrix is constructed by

$$\boldsymbol{H}^a = \left[ \boldsymbol{H}_y \quad \boldsymbol{H}_b \quad \boldsymbol{H}_d \right] \tag{76}$$

The matrix $\boldsymbol{H}_y$ is calculated using the state transmission matrix and the gravity gradient Jacobian matrix. The matrix $\boldsymbol{H}_b$ is defined in Eq. (68) and $\boldsymbol{H}_d$ is defined as follows

$$\boldsymbol{H}_d = \begin{bmatrix} \boldsymbol{0}_{6\times6} \\ (t_1 - t_0)\boldsymbol{I}_{6\times6} \\ \vdots \\ \left(t_{N_j-1} - t_0\right)\boldsymbol{I}_{6\times6} \end{bmatrix}_{(6N_j)\times6} \tag{77}$$

The state update and the final covariance are the same as Eq. (40) and Eq.(54), respectively.

## 5. Results

The orbit determination algorithm has been tested with the actual GOCE data. The Level 1b product EGG_NOM_1b which contains the raw GGT measurements (EGG_GGT dataset) and the gradiometer inertial attitude quaternions (EGG_IAQ dataset) are used as inputs [35]. The precise orbit solutions provided in the Level 2 product SST_PSO_2 (reduced-dynamic orbits from GPS) are used to evaluate the accuracy of the GGT derived orbits [44]. The test covers 12-hour arc data starting from 8 September 2013, 00:00:00.0 (GPS Time). The data are reported to have good quality and no special events (data anomaly or calibration) occurred. The data are resampled at intervals of 30s. Thus the gravity gradients accumulate to a total number of 8640 measurements.

The position and velocity as well as the biases are estimated using the ILS filter, the ULS filter, and the adaptive HLS filter, respectively. The initial errors added in position and velocity are set to [$10^4$ m, $10^4$ m, $10^4$ m, 10 m/s, 10 m/s, 10 m/s] and the diagonal elements of the initial covariance of position and velocity are set to [($10^4$ m)$^2$, ($10^4$ m)$^2$, ($10^4$ m)$^2$, (10 m/s)$^2$, (10 m/s)$^2$, (10 m/s)$^2$]. The initial errors added in the 6 biases are set to [10 E, 10 E, 10 E, 10 E, 10 E, 10 E] and the diagonal elements of the initial covariance of the biases are set to [(10 E)$^2$, (10 E)$^2$, (10 E)$^2$, (10 E)$^2$, (10 E)$^2$, (10 E)$^2$]. The bias drifts are all assumed to be zero. The relative error tolerance for the convergence criterion is set to $10^{-5}$ and the maximum iteration number is set to 10. The fudge factors of the covariance computation are all set to 10. Other values of the fudge factors can also be used. For the ULS filter and the adaptive HLS filter, the three scaling parameters of the unscented transformation, i.e., $\alpha$, $\beta$, and $\kappa$ are set to 1, 2, and 0, respectively. The tolerance of the position uncertainty in the adaptive HLS filter is set to $5\times10^4$ m.

Figs. 4(a) and 4(b) show the histories of iterations and the variations of the position and velocity errors. The ULS filter and the adaptive HLS filter are successfully converged, whereas the ILS filter fails to converge due to the large initial errors. The adaptive HLS filter shows better convergence performance and better accuracy than the ULS filter since the system's nonlinearity are kept much lower by the adaptive adjustment of measurement time span and the iterative updating of the covariance matrix. Another advantage of the adaptive HLS filter is the short computation time. For each iteration loop, the execution time of the ULS filter is 8107 s and the execution time of the adaptive HLS filter is 2382 s. The reason is that the ULS filter propagates 13 (= $2L + 1$) sigma points of orbits at each time, whereas the adaptive HLS filter propagates one orbit only.

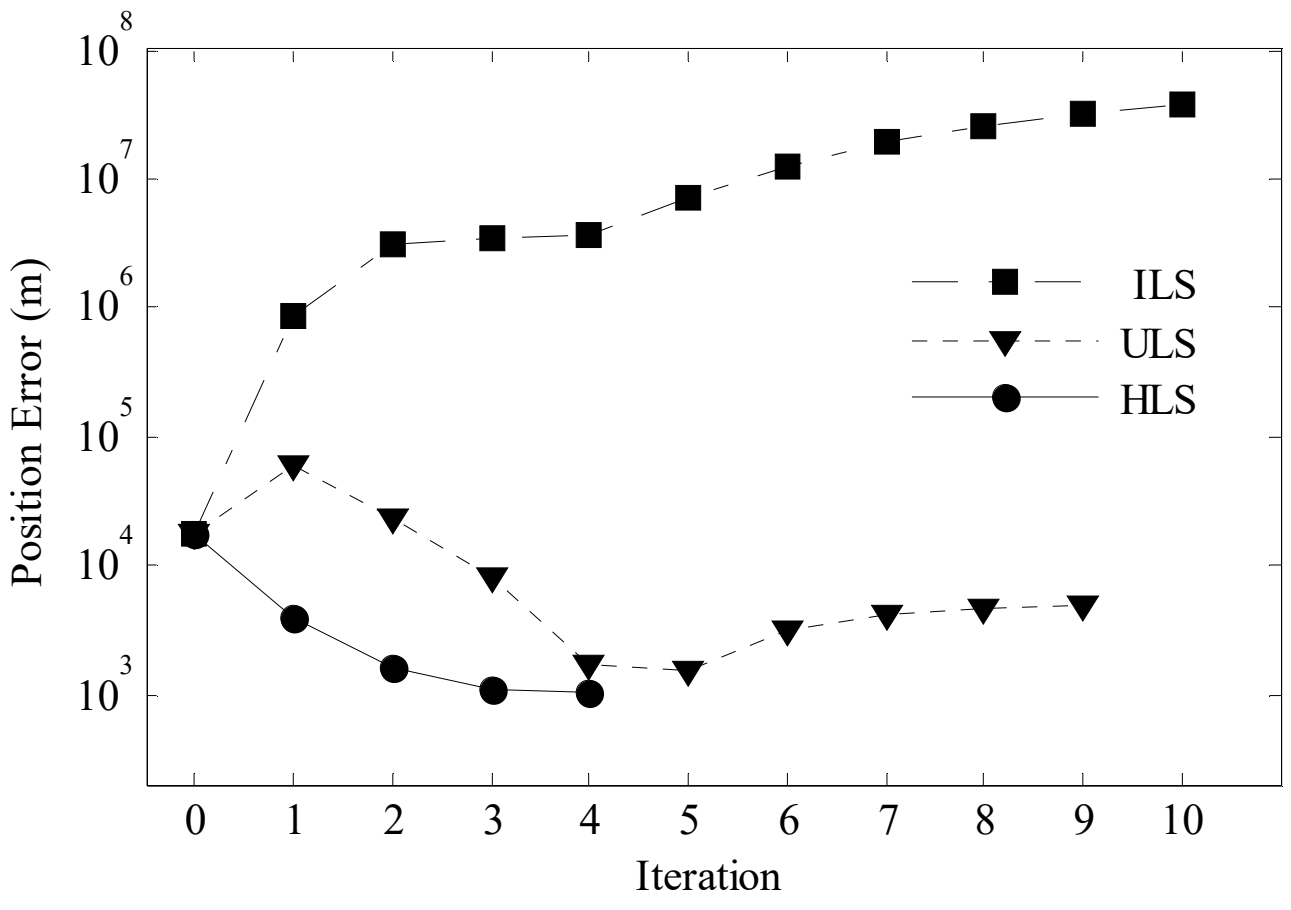

(a) Position errors

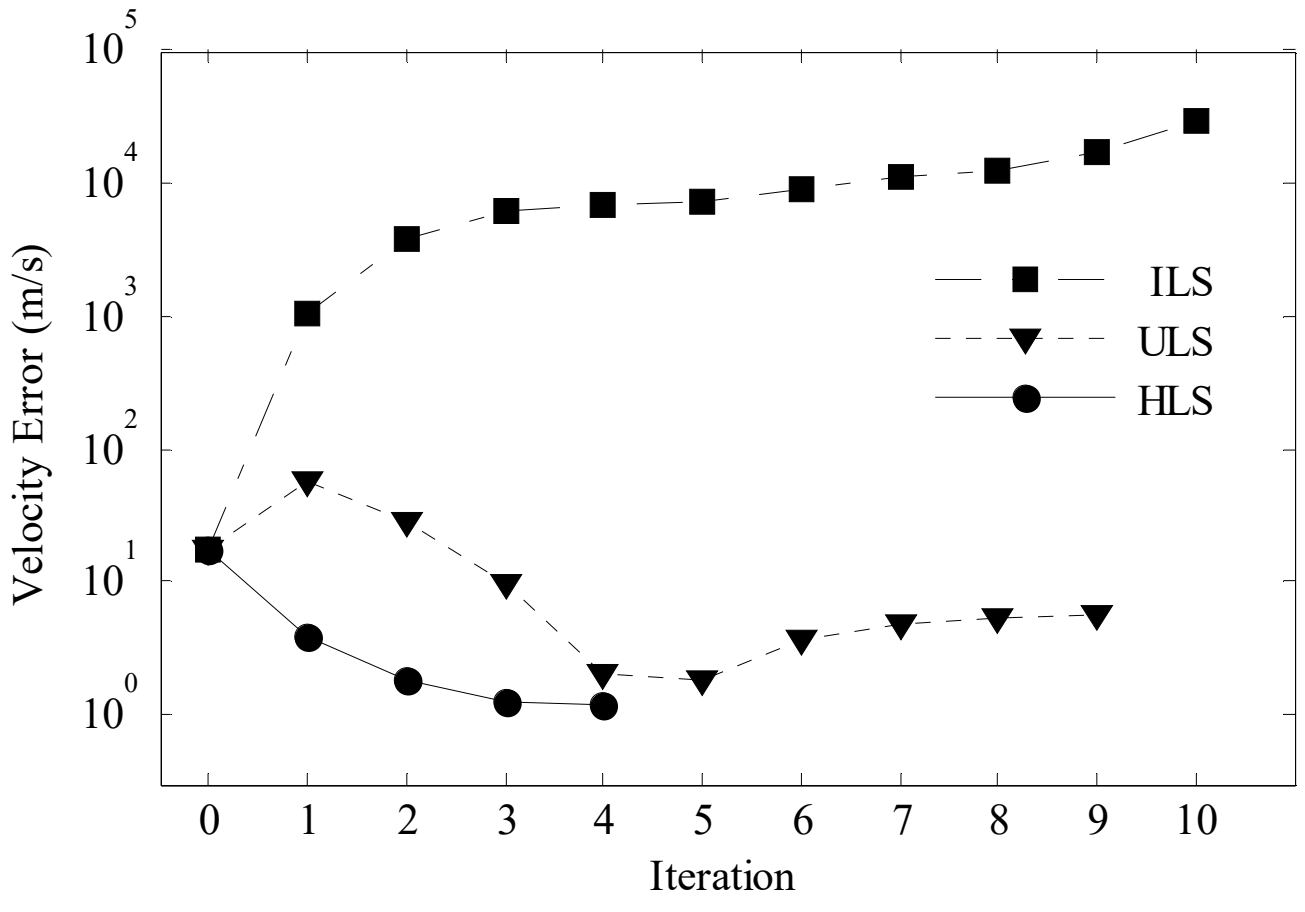

(b) Velocity errors

**Fig. 4**. Histories of iterations and the variations of the position and velocity errors of the ILS, the ULS, and the adaptive HLS.

The estimated initial position and velocity from the adaptive HLS filter have been used to generate the orbit ephemeris which is compared with the GPS-derived orbit trajectory. The position and velocity errors (radial, along-track, and cross-track components) varying with time are shown in Figs. 5(a) and 5(b). The root mean square (RMS) values of the position errors are 10.8 m, 1208.3 m, and 37.9 m, respectively. The RMS values of the velocity errors are 1.2 m/s, 0.013 m/s and 0.044 m/s, respectively. The large along-track position error (negative) and the large radial velocity error (positive) indicate that the orbit estimated from the adaptive HLS filter is a trailing orbit of the true orbit. This phenomenon is due to the estimation errors of the biases.

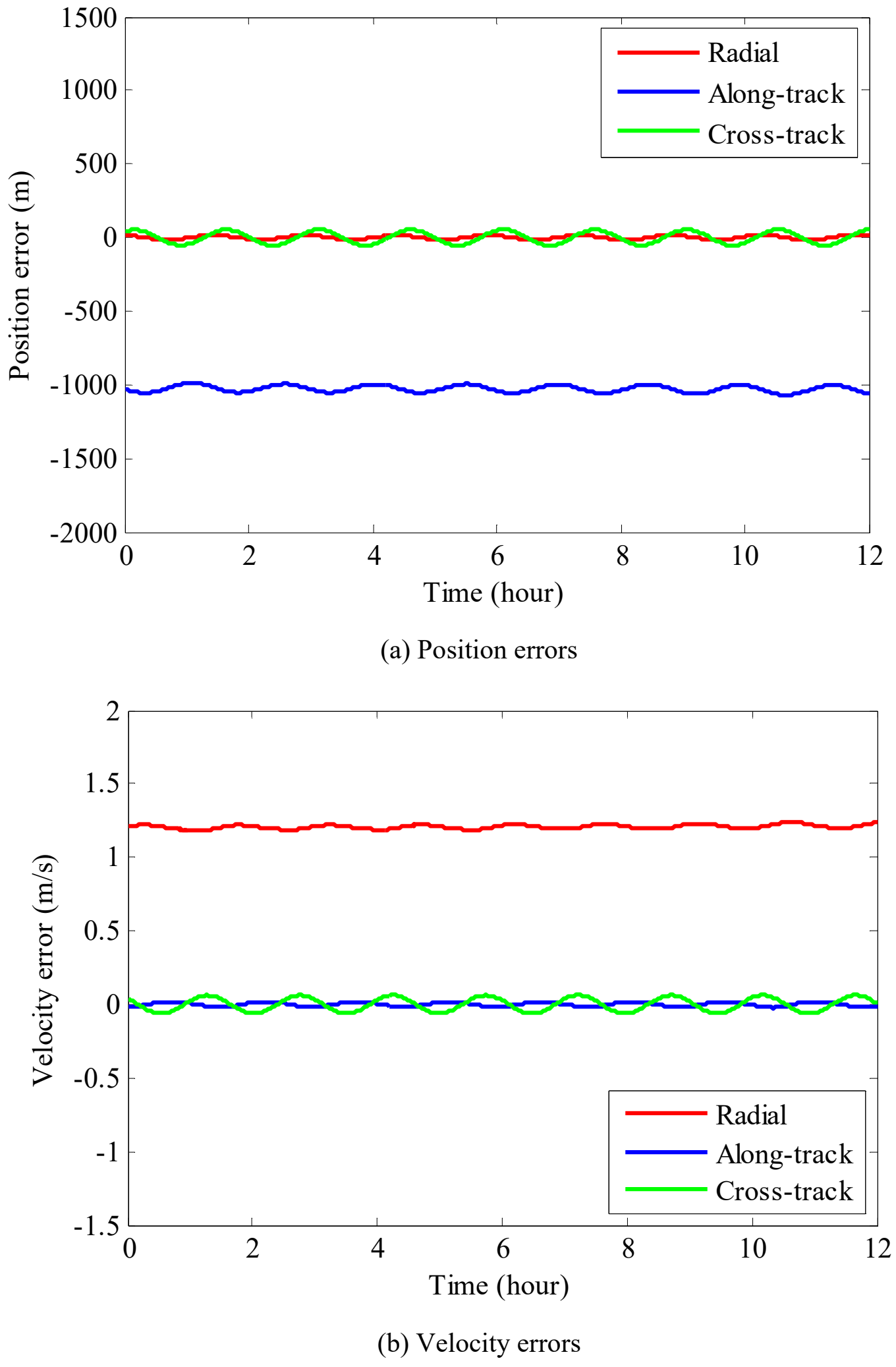

**Fig. 5**. Evolution of the radial, along-track, and cross-track position and velocity errors of the adaptive HLS filter.

Table 1 lists the reference values of the biases and the drifts, which are determined from the differences between the actual measurements and the gravity gradients computed using the GPS-derived orbits and a 300×300 subset of the EGM2008 gravity model. The estimation errors of the 6 biases are -3.7 mE, 64.3 mE, 0.4 mE, -542 mE, -641.9 mE, and -314.0 mE, respectively. The poor observability of the $b_{xz}$ component results in the biases in the orbit estimation. The post-fit measurement residuals of the adaptive HLS filter are also investigated and are illustrated in Fig. 6. The obvious drifts in the residuals (*yy*, *xy*, and *yz* components) are due to the underfitting of the filter.

**Table 1**. Reference values of the initial biases and the drifts

| Component | Initial bias, E | Drift, mE/h |
|---|---|---|
| *xx* | 532.71 | -1.10 |
| *yy* | -755.26 | 11.53 |
| *zz* | -217.51 | 0.11 |
| *xy* | -5909.76 | -89.75 |
| *xz* | -26.60 | 0.28 |
| *yz* | -3171.07 | 47.30 |

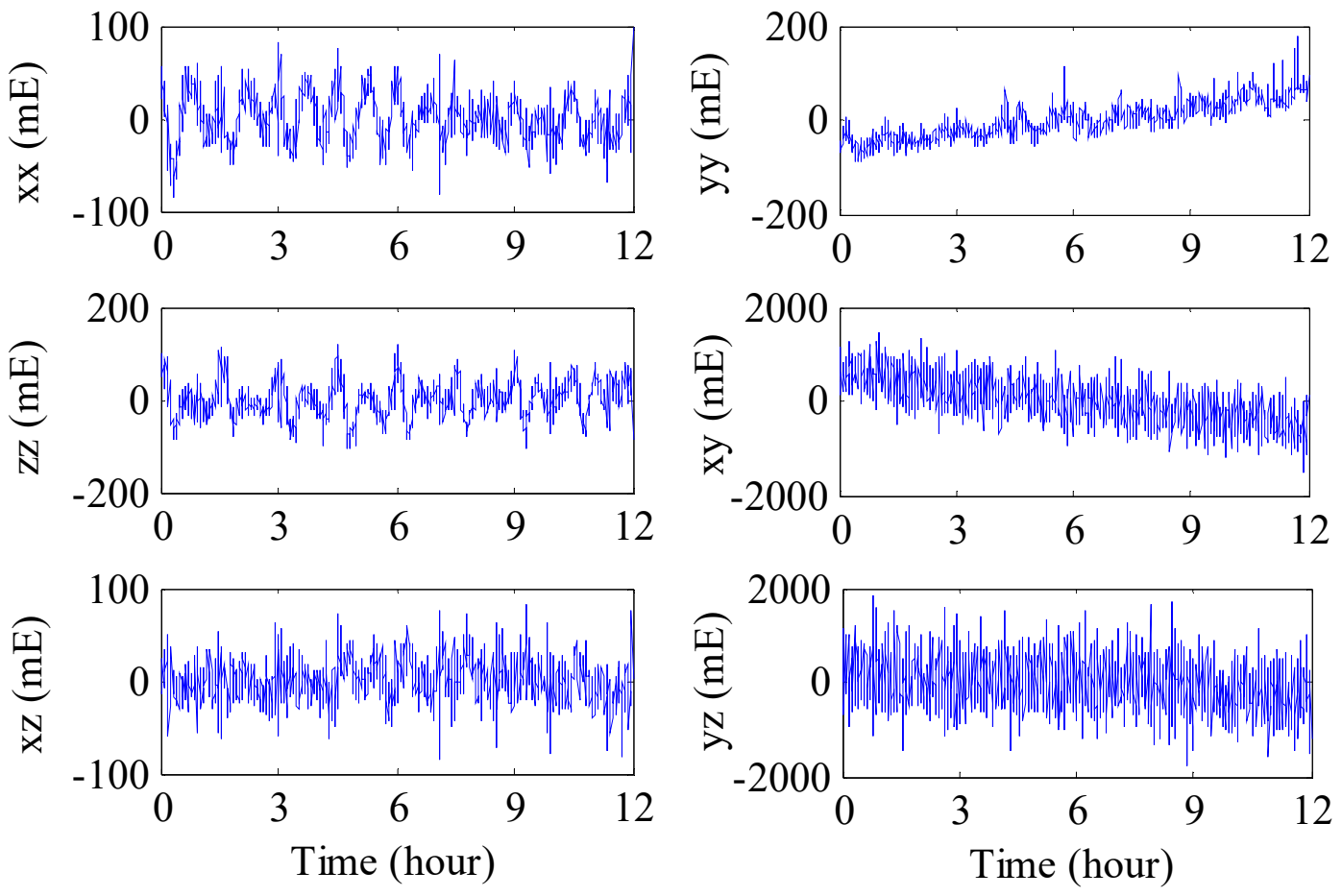

**Fig. 6**. Post-fit measurement residuals of the adaptive HLS filter.

The bias drifts are further estimated using the augmented state ILS filter. The initial values of the orbital states and the biases are set to the estimates obtained from the adaptive HLS filter. The initial covariance of these parameters is set to be the same as that for the adaptive HLS filter. The initial values of the drifts are all set to be zero. The diagonal elements of the initial covariance of the drifts are set to [(1 mE/h)$^2$, (10 mE/h)$^2$, (1 mE/h)$^2$, (100 mE/h)$^2$, (1 mE/h)$^2$, (100 mE/h)$^2$]. The augmented state ILS filter successfully estimates the drifts and improves the orbit determination accuracy. Figs 7(a) and 7(b) plot the position and velocity errors varying with time after the drift estimation. The RMS values of the position errors are reduced to 10.4 m, 677.0 m, and 22.8 m. The RMS values of the velocity errors are reduced to 0.80 m/s, 0.012 m/s and 0.026 m/s. The estimation errors of the 6 biases are reduced to -1.32 mE, 7.65 mE, -4.63 mE, 22.40 mE, -420.31 mE, and -44.59 mE, respectively. The estimation errors of the drifts are 0.59 mE/h, -2.07 mE/h, 0.84 mE/h, -4.77 mE/h, 0.21 mE/h, and 2.76 mE/h, respectively. The post-fit measurement residuals after drift estimation are shown in Fig. 8. It is seen that there are no drifts in the residuals. For the ultra-sensitive components (*xx*, *yy*, *zz*, and *xz*) the residuals are dominated by the low-frequency noises. For the less-sensitive components (*xy* and *yz*) white noises dominate the residuals.

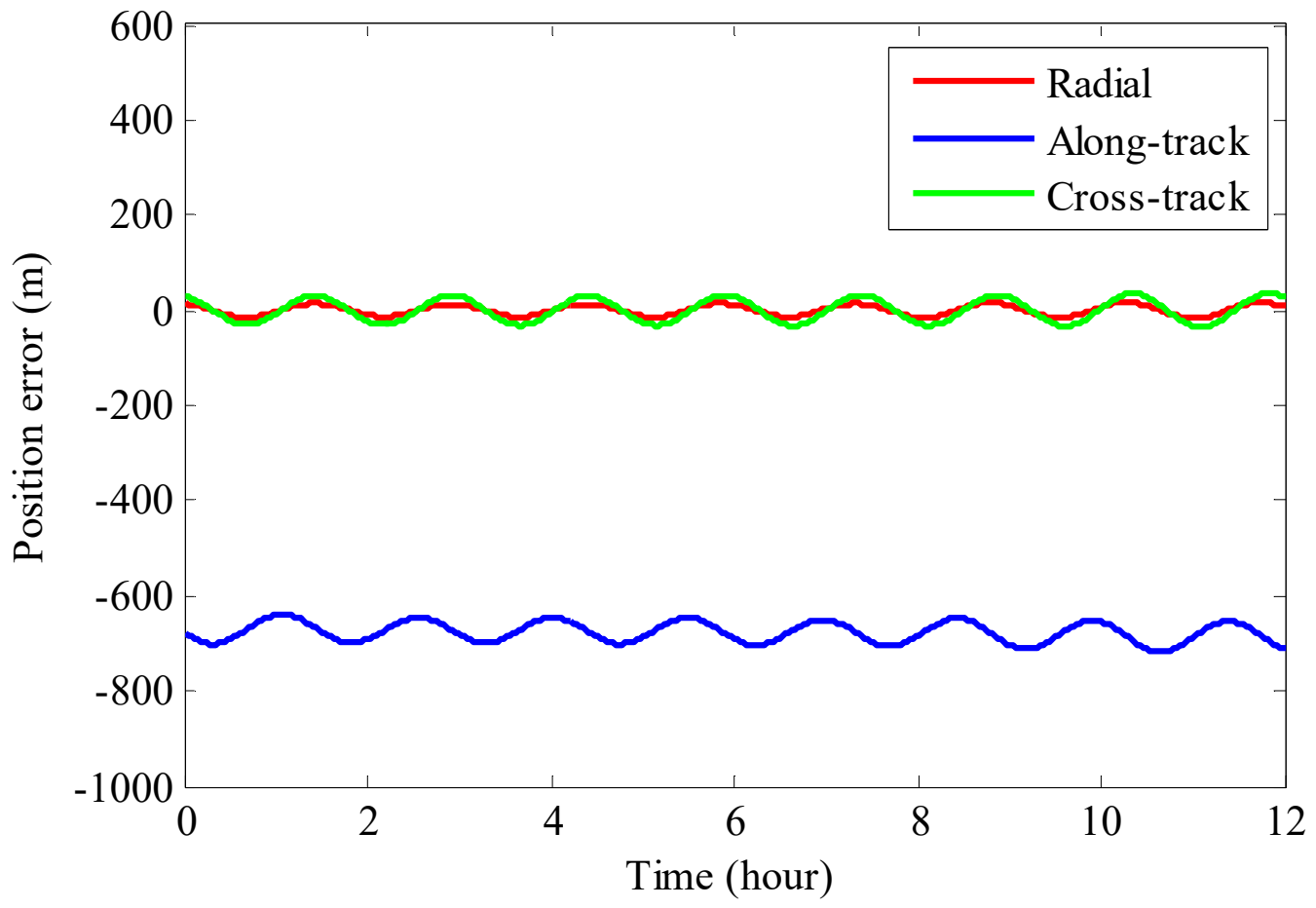

(a) Position errors

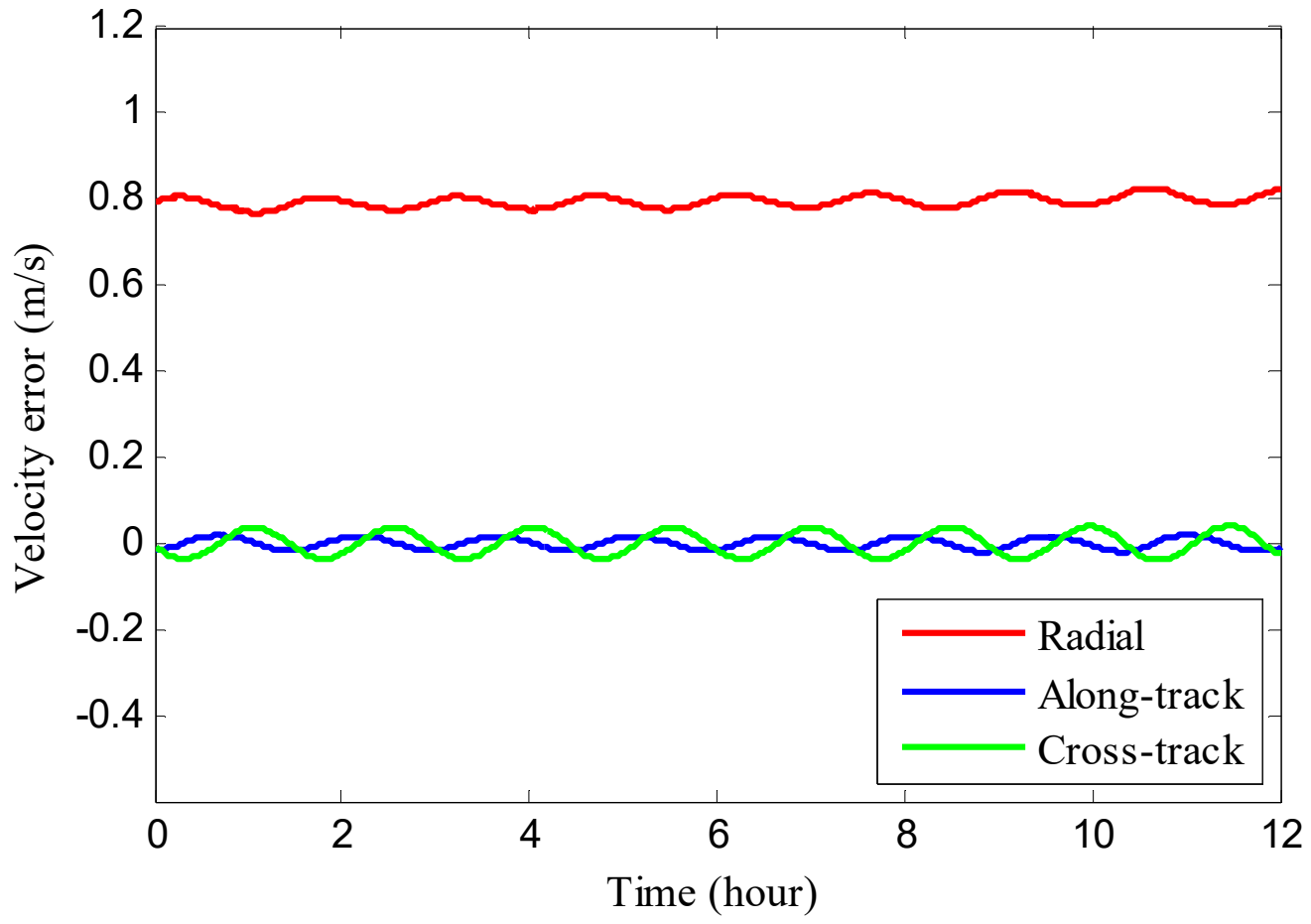

(b) Velocity errors

**Fig. 7**. Evolution of the radial, along-track, and cross-track position and velocity errors after the drift estimation.

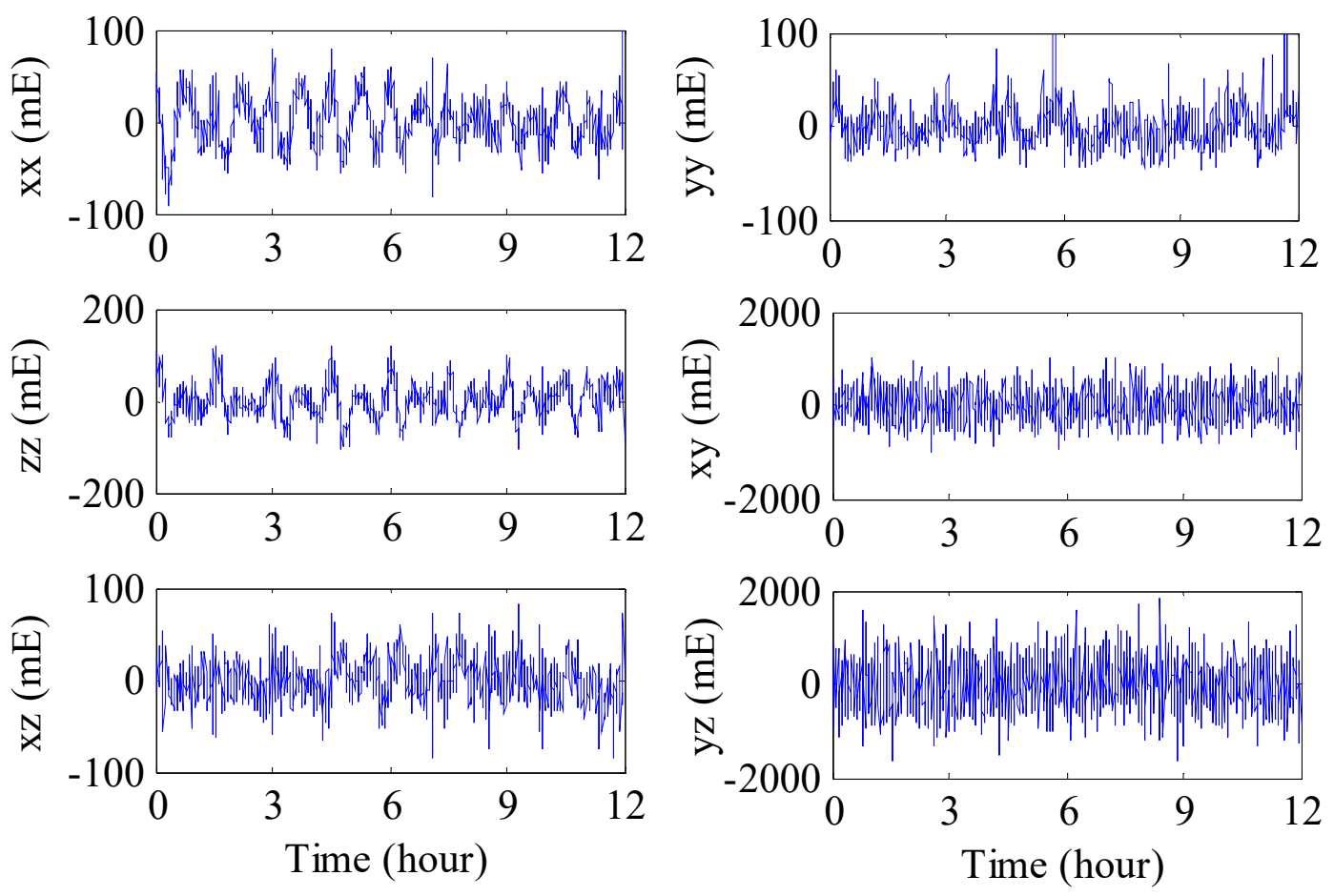

**Fig. 8**. Post-fit measurement residuals after drift estimation.

Besides the EGM2008 model, four other global gravity models including EGM96 [45], JGM3 [46], EIGEN-6C4 [47], and GGM05G [48] have also been tested to provide a comparative analysis. The EGM96 was developed jointly by NGA (formerly known as NIMA), NASA Goddard and Ohio State University and is complete to degree and order 360, whereas the JGM3 describes the Earth gravity field up to degree and order 70, developed by the University of Texas and NASA/GSFC. The EIGEN-6C4 model was jointly developed by GFZ Potsdam and GRGS Toulouse and is computed from a combination of LAGEOS, GRACE, GOCE and surface gravity data. The GGM05G model is an unconstrained global gravity model complete to degree and order 240 and is determined from the GRACE and GOCE data. The stochastic RMS orbit differences of these four models with respect to the solutions obtained from the EGM2008 model are listed in Table 2. The EGM96, EIGEN-6C4, and GGM05G gravity models show good consistency with the EGM2008 model for the gravity gradient based orbit determination. Slightly larger discrepancies in the along-track position (56.6 m) and the radial velocity (0.0663 m/s) components are found between JGM3 and EGM2008. Nevertheless, the discrepancies are small compared to the absolute orbit determination errors.

**Table 2**. RMS orbit differences with respect to EGM2008 for the EGM96, JGM3, EIGEN-6C4, and GGM05G models

| Gravity model | Position difference, m | | | Velocity difference, m/s | | |
|---|---|---|---|---|---|---|
| | Radial | Along-track | Cross-track | Radial | Along-track | Cross-track |
| EGM96 | 1.51 | 3.47 | 2.32 | $2.89\times10^{-3}$ | $1.77\times10^{-3}$ | $2.74\times10^{-3}$ |
| JGM3 | 1.66 | 56.6 | 6.92 | $6.63\times10^{-2}$ | $1.93\times10^{-3}$ | $8.03\times10^{-3}$ |
| EIGEN-6C4 | $4.33\times10^{-2}$ | 13.05 | 0.57 | $1.53\times10^{-2}$ | $6.04\times10^{-5}$ | $6.55\times10^{-4}$ |
| GGM05G | $5.24\times10^{-2}$ | 12.15 | 0.55 | $1.43\times10^{-2}$ | $7.38\times10^{-8}$ | $6.30\times10^{-4}$ |

## 6. Conclusions

In this paper, an orbit determination method using gravity gradient measurements has been described and the GOCE satellite has been used as a case study. Within the strategy, satellite orbits are estimated from a combined usage of the spaceborne gravity gradiometer and star trackers. Actually, the function of the star trackers is two-fold: to provide estimations of angular rates and angular accelerations for gravity gradients retrieval and to provide precise attitude quaternions in order to isolate the orientation contributions.

The orbit determination is implemented by an adaptive hybrid least squares batch filter. The performance of the algorithm is evaluated using the actual GOCE data, and a position accuracy of tens of meters has been achieved for the radial and cross-track position components. The large along-track position error is due to the poor observation of one of the measurement biases. The low-frequency noises remain in the measurement residuals and need to be dealt with in the future study to further improve the accuracy. Nevertheless, the present work demonstrates the feasibility of orbit determination from gravity gradients containing drifting biases and provides an alternative autonomous navigation method for satellites in near-Earth orbits.

## Acknowledgements

This research was supported by the National Natural Science Foundation of China (No. 11002008). The European Space Agency is acknowledged for providing the GOCE data. The authors acknowledge the valuable remarks of the reviewers and the editor.